\def\xflux{~erg~cm$^{-2}$~s$^{-1}$}
\def\xlum{~erg~s$^{-1}$}

\documentclass[preprint]{aastex}
\usepackage{emulateapj5,apjfonts}
\tighten

\journalinfo{The Astronomical Journal, 2002 March, astro-ph/0112002}
\slugcomment{Received 2001 September 25; accepted 2001 November 29}
\shortauthors{BAUER ET AL.}
\shorttitle{{\it CHANDRA} DEEP FIELD NORTH EXTENDED X-RAY SOURCES}

\begin{document}

\title{The {\it Chandra} Deep Field North Survey. IX. Extended X-ray Sources\altaffilmark{1}}

\author{
F.~E.~Bauer,\altaffilmark{2} D.~M.~Alexander,\altaffilmark{2}
W.~N.~Brandt,\altaffilmark{2} A.~E.~Hornschemeier,\altaffilmark{2}
T.~Miyaji,\altaffilmark{3}
G.~P.~Garmire,\altaffilmark{2} D.~P.~Schneider,\altaffilmark{2}
M.~W.~Bautz,\altaffilmark{4}
G.~Chartas,\altaffilmark{2}
R.~E.~Griffiths\altaffilmark{3},
and W.L.W.~Sargent\altaffilmark{5}
}

\altaffiltext{1}{Based on observations obtained with the Hobby-Eberly Telescope, 
which is a joint project of the University of Texas at Austin, the
Pennsylvania State University, Stanford University,
Ludwig-Maximillians-Universität München, and Georg-August-Universität
Göttingen.}

\altaffiltext{2}{Department of Astronomy and Astrophysics, 525 Davey Lab, 
The Pennsylvania State University, University Park, PA 16802.}

\altaffiltext{3}{Department of Physics, Carnegie Mellon University, Pittsburgh, PA 15213.}

\altaffiltext{4}{Massachusetts Institute of Technology, Center for Space Research,
70 Vassar Street, Building 37, Cambridge, MA 02139.}

\altaffiltext{5}{Palomar Observatory, California Institute of Technology, Pasadena, CA 91125.}

\begin{abstract}
The $\approx$~1~Ms {\it Chandra} Deep Field North observation is used
to study the extended X-ray sources in the region surrounding the
Hubble Deep Field North (HDF-N), yielding the most sensitive probe of
extended X-ray emission at cosmological distances to date. A total of
six such sources are detected, the majority of which align with small
numbers of optically bright galaxies. Their angular sizes, band
ratios, and X-ray luminosities --- assuming they lie at the same
distances as the galaxies coincident with the X-ray emission --- are
generally consistent with the properties found for nearby groups of
galaxies. One source is notably different and is likely to be a
poor-to-moderate X-ray cluster at high redshift (i.e., $z\ga0.7$). 
This source has a large angular extent, a double-peaked X-ray
morphology, and an overdensity of unusual objects [Very Red Objects,
optically faint ($I\ge24$) radio and X-ray sources]. Another of the
six is coincident with several $z\approx1.01$ galaxies located within
the HDF-N itself, including the FR~I radio galaxy
VLA~J123644.3+621133, and is likely to be a group or poor cluster of
galaxies at that redshift. We are also able to place strong
constraints on the optically detected cluster of galaxies
ClG~1236$+$6215 at $z=0.85$ and the wide-angle-tail radio galaxy
VLA~J123725.7$+$621128 at $z\sim1$--2; both sources are expected to
have considerable associated diffuse X-ray emission, and yet they have
rest-frame 0.5--2.0~keV X-ray luminosities of $\la3\times10^{42}$\xlum
and $\la(3$--$15)\times10^{42}$\xlum, respectively. The environments
of both sources are either likely to have a significant deficit of hot
intra-cluster gas compared to local clusters of galaxies, or they are
X-ray groups. We find the surface density of extended X-ray sources in
this observation to be 167$^{+97}_{-67}$ deg$^{-2}$ at a limiting
soft-band flux of $\approx3\times10^{-16}$\xflux. No evolution in
the X-ray luminosity function of clusters is needed to explain this value.

\end{abstract}


\section{INTRODUCTION}\label{intro}

The Hubble Deep Field North
\citep[HDF-N;][]{Williams1996,Ferguson2000} was chosen as the location
of a deep {\it Hubble Space Telescope} survey because it contained no
known bright sources at radio, infrared, optical, or X-ray wavelengths
and no known nearby ($z<0.3$) clusters of galaxies. This effort was
conceived to advance the study of galaxy evolution to high redshifts,
but it has since initiated one of the most intensive, multi-wavelength
investigations on the sky, influencing a wide range of astronomical
topics (e.g., Ferguson et al. 2000 and references therein) and
yielding one of the most comprehensive data sets publicly available
(e.g., deep imaging at nearly all astronomically accessible
wavelengths and more than 700 spectroscopic redshifts within a
$\sim4\arcmin$ radius of the HDF-N).\footnote{See the Hubble Deep
Field Clearinghouse;
http://www.stsci.edu/ftp/science/hdf/clearinghouse/clearinghouse.html.} 
Recently, the {\it Chandra X-ray Observatory} \citep[hereafter {\it
Chandra};][]{Weisskopf2000} completed an $\approx$~1~Ms survey of the
HDF-N and its environs, providing an extremely sensitive view of the
X-ray Universe.

The {\it Chandra} Deep Field North Survey \citep[CDF-N;][hereafter
Paper~V]{Brandt2001b} covers an area approximately $18\arcmin \times
22\arcmin$ in size and reaches 0.5--2.0~keV (soft) and 2.0--8.0~keV
(hard) flux limits of $\approx 3\times 10^{-17}$\xflux\ and $\approx
2\times 10^{-16}$\xflux\ (5.5$\sigma$) near the aimpoint for point sources. In
addition to resolving most of the X-ray background into individual
point sources \citep[e.g., Paper V,][]{Cowie2002}, the
observation allows the detection of even relatively poor clusters and
groups of galaxies to significant redshifts ($z\sim1$). The
detection of extended X-ray emission from hot gas provides compelling
evidence that apparent optical clusters or groups are truly
gravitationally bound. Since the properties of clusters and groups are
intimately dependent upon cosmological parameters and the history of
structure formation, X-ray emission offers a useful probe of
hierarchical structure.

There is already substantial observational evidence that sources in
the vicinity of the HDF-N tend to cluster at certain redshifts
in ``walls'' or ``filaments,'' and that a few of these redshift peaks
can be broken up spatially into apparent groups of galaxies
\citep[e.g.,][]{Cohen2000, Dawson2001}. The CDF-N has the potential to
determine whether the gravitational potential wells of these apparent
groups are deep enough to harbor large amounts of hot gas and dark
matter. Additionally, deep radio imaging of this region has revealed
two highly extended radio sources \citep{Richards1998, Muxlow1999,
Snellen2001}; one is the Fanaroff-Riley~I
\citep[FR~I;][]{Fanaroff1974} radio galaxy VLA~J123644.3+621133
located within the HDF-N itself, and the other is the wide-angle-tail
(WAT) source VLA~J123725.7$+$621128. These two radio 
\end{multicols}
\vbox{
\vspace{17.3cm}
\vspace{0.0cm} 
\figcaption[fig1.color.eps]
{Adaptively smoothed ``true color'' X-ray image with red, green, and
blue representing 0.5--2.0~keV, 2.0--4.0~keV, and 4.0--8.0~keV
emission, respectively. Prior to combination, each color image was
smoothed to 2.5$\sigma$ with an adaptive kernel algorithm
\citep{Ebeling2001} to enhance diffuse features. The white boxes
denote the six extended X-ray sources detected (see $\S$\ref{notes}),
while the two dashed circles indicate cluster candidates found at
other wavelengths that show no significant extended X-ray emission (see
$\S$\ref{nondetections}). In order to show clearly the extended
sources, it was necessary to saturate bright X-ray point sources. This
tends to make all of the brighter point sources appear yellow in the
image, regardless of whether their spectral shapes are particularly
hard or soft.\label{fig:xrayfig}}
}
\begin{multicols}{2}

\noindent
sources are
notable because FR~Is, and in particular WATs, are known to reside
predominantly in or near rich clusters of galaxies; extended X-ray
emission associated with these particular radio sources has yet to be
detected. Arguably the most definitive evidence for clustering near
the HDF-N, however, comes from \citet{Dawson2001} who recently
reported the discovery of the $z=0.85$ cluster of galaxies ClG
1236$+$6215 slightly north of the HDF-N. Dawson et al. predict this
cluster should have a bolometric X-ray luminosity in excess of
$10^{44}$\xlum, which, for an extended object of radius 30$\arcsec$ at
a redshift of $z=0.85$, should be nearly a factor of 15 above the
detection threshold of the CDF-N. Although a 21~ks {\it ROSAT} High
Resolution Imager (HRI) observation of this region detected no
extended X-ray objects, the resulting soft-band flux threshold of
$\sim$2$\times10^{-14}$\xflux\ is not particularly constraining; for
instance, the HRI observation would have missed a typical
$10^{43}$\xlum\ X-ray cluster at $z\ga0.4$ or even a typical
$10^{44}$\xlum\ X-ray cluster at $z\ga0.95$. The 1~Ms {\it Chandra}
observation allows us to push these constraints much lower.

The deepest soft-band X-ray observations prior to {\it Chandra} were
those of the {\it ROSAT} Ultra Deep Survey toward the Lockman Hole
\citep[hereafter UDS; e.g.,][]{Lehmann2001}, which detected ten extended
objects over a $\sim$$30\arcmin$ field of view. A few of these sources
are classified as clusters \citep[including the double-peaked
lensing cluster RX~J105343$+$5735;][]{Hasinger1998}, but the majority
appear to be groups; all are thought to lie at redshifts of
$z\sim0.2$--1.0. Results from other deep surveys with {\it ROSAT}
\citep[e.g.,][]{McHardy1998, Zamorani1999} were comparable. 
While the CDF-N covers only $\approx$~$\onequarter$ the area of the
UDS, it probes the X-ray sky a factor of $\approx~7$ times deeper for
extended objects in the soft band, making the discovery of fainter and
more distant objects possible.

Here we describe the nature of the extended X-ray sources detected
within the extremely deep CDF-N observation. In $\S$\ref{data} and
$\S$\ref{optical}, we briefly outline our reduction and detection
techniques for the X-ray and optical observations, respectively. 
Descriptions of individual sources are presented in $\S$\ref{notes}. 
Finally our findings are discussed and summarized in
$\S$\ref{discuss}. Throughout this paper, we adopt
$H_{0}=65$~km~s$^{-1}$~Mpc$^{-1}$, $\Omega_{\rm M}=\onethird$, and
$\Omega_{\Lambda}=\twothirds$. The Galactic column~density toward the
CDF-N is $(1.6\pm0.4)\times10^{20}$ cm$^{-2}$
\citep{Stark1992}. Coordinates are for the J2000 epoch.

\section{Chandra Observations and Reduction Techniques}\label{data}

A full description of the CDF-N observations, data reduction methods,
and catalog of detected sources is provided elsewhere (Paper~V). Here
we outline only the methods relevant to detect extended X-ray sources
in the CDF-N and measure their characteristics. The reduction and
analysis detailed below were carried out using the {\it Chandra}
Interactive Analysis of Observations ({\sc ciao}) Version~2.1 tools
whenever possible,\footnote{See http://asc.harvard.edu/ciao/.} 
although custom software was sometimes also required.

\subsection{Source Detection}\label{detect_procedure}

Extended source searching was performed over the entire ACIS-I field
using the Voronoi Tessellation and Percolation algorithm
\citep[VTP;][]{Ebeling1993, Dobrzycki1999}. Specifically, we created soft
(0.5--2.0~keV), hard (2.0--8.0~keV) and full (0.5--8.0~keV) band
images using the standard {\it ASCA} grade set\footnote{Identical
analysis was carried out on event lists and images made using the
restricted ACIS grade set, since this grade set has been shown to
reduce the particle background by up to $\approx30\%$ \citep[see
Table~1 of [hereafter Paper~IV]{Brandt2001a}. The results were
quantitatively similar to those found with the {\it ASCA} grade set
and did not greatly enhance the signal-to-noise of any source.} 
defined in Table~2 of Paper~V. These images were then searched for
extended sources using the {\sc ciao} tool {\sc vtpdetect} adopting a
false-positive probability threshold of $1\times10^{-7}$ and a
``coarse'' parameter of 100. We required that {\sc vtpdetect}-detected
extended sources satisfy each of the following criteria: (1) average
{\sc vtpdetect} radii\footnote{Defined as the average of the major and
minor axes reported by {\sc vtpdetect}.} greater than three times the
95\% encircled-energy radius of the point spread function (PSF) at a
given off-axis angle, (2) visible extended X-ray emission in
exposure-corrected, adaptively smoothed images \citep[made with the
{\sc ciao} tool {\sc csmooth}; see][]{Ebeling2001}, and (3) a
signal-to-noise of $>$~3 above the local background derived via
aperture photometry (see $\S$\ref{properties}). While we appreciate
that the $3\sigma$ criteria is less stringent than the
$1\times10^{-7}$ false-positive probability threshold of {\sc
vtpdetect}, it allowed us to remove manually point sources that might
have subtly affected the {\sc vtpdetect} search algorithm and gives us
a secondary significance estimator. In total, six extended X-ray
sources were considered to be legitimate. Two of these are solid
detections while the other four all lie close to the $3\sigma$ limit. 
None of these lies directly on ACIS-I chip gaps or edges, and the
exposure maps are relatively smooth near all of the detected sources
(with variations of $\sim$20\% at most). This suggests that the data
are less sensitive to extended emission on chip gaps or edges. Note
that all of the sources lie within the ``high exposure area'' where
the median effective exposure time is $>$~800~ks (see Figure 7 of
Paper~V).

Figure~\ref{fig:xrayfig} shows a ``true color'' image of the CDF-N,
with the colors red, green, and blue representing the 0.5--2.0~keV,
2.0--4.0~keV, and 4.0--8.0~keV energy bands, respectively. Each X-ray
image was smoothed with {\sc csmooth} prior to combination to permit
the simultaneous viewing of compact and extended sources. The smoothed
images each have a signal-to-noise ratio of 2.5 per smoothing beam. 
The six extended X-ray sources are labeled and enclosed within boxes
corresponding to the size of the optical cut-out images described in
$\S$\ref{optical}. Also shown are two cluster candidates not detected
in this observation (see $\S$\ref{nondetections} for details).

The CDF-N observations spanned approximately 16 months and were taken
at a variety of roll angles (see Table~1 of Paper~V), so the summed
diffuse particle background is both temporally and spatially
variable.\footnote{For a detailed characterization of {\it Chandra's}
background, see http://asc.harvard.edu/cal/Links/Acis/acis/.} This
non-uniformity hinders an accurate measure of the local X-ray
background and makes determination of sample completeness difficult. 
Furthermore, it is possible that the backgrounds from individual
observations could combine together in such a manner that local
background enhancements by a factor of a few are common. If such
enhancements exist in the CDF-N, they could in turn lead to spurious
detections. We can check for such problems since instrumental
background features are likely to occur at the same position on the
CCD regardless of aim point or roll angle while cosmic sources will
not. We therefore split the $\approx$~1~Ms observation into two
adaptively smoothed soft-band images of 432~ks and 543~ks, made using
only data with roll angles of 36\fdg4--44\fdg5 and 134\fdg3--143\fdg8,
respectively (see Table~1 of Paper~V). The six extended sources
mentioned above were the only regions of diffuse emission visible in
both images, arguing against an instrumental origin.

Finally, because of the nature of the CDF-N observations noted above,
we caution that any extended X-ray sources that happen either to
intersect one of the CCD chip gaps or fall along the edges of
observations could potentially have irregular morphologies. This could
in turn lead to an inaccurate assessment of their physical nature. The
fact that none of the CDF-N extended sources detected in our analysis
lies on CCD chip gaps or edges is therefore extremely important, as it
implies that the peculiar morphological features seen in the brightest two
extended X-ray sources in Figure~\ref{fig:xrayfig} are likely to be
real.

\subsection{Source Properties}\label{properties}

The basic properties of the six extended CDF-N sources are given in
Table~\ref{tab:sources}. Since sources were best detected in the
soft band, the poorly defined source positions were estimated by eye
from the adaptively smoothed soft-band image. The counts for these
sources were determined via manual aperture photometry. The sizes and
shapes of the apertures were chosen to encompass the apparent X-ray
emission associated with the sources as determined from the adaptively
smoothed images; column~3 of Table~\ref{tab:sources} lists these
regions. Background counts were extracted from annular regions
immediately outside the source extraction regions. Point sources and
regions of strongly varying background were excluded. The vignetting
correction applied to the flux calculation below was determined by
extracting average exposure times in the exposure map from source and
background regions identical to those used to extract the counts. The
count rates associated with these extended sources are typically only
a fraction of the area-corrected, point-source excluded, average
X-ray background rate. The uncertainties in the measured source counts
are therefore large and increase with the extraction region size. 
Source-to-background (S-to-B) ratios for each source are listed in
column~6.

For all but two sources (1 and 4), the numbers of counts in the
diffuse emission are comparable to or larger than those from nearby
point sources. We would not expect this many counts from the extended
wings of the PSF of an isolated bright point source alone, and
therefore the counts must be produced by these extended sources. In
the cases of sources~1 and 4, even though the counts from the
background-subtracted diffuse emission are a factor of 10--20 less than
the total point-source counts, the centroids of the diffuse emission
are distinctly offset from bright point sources by
$\approx$~15$\arcsec$. Such asymmetric offsets are unlikely to be due
to the PSF wings of bright embedded point sources.

All of the extended sources in Figure~\ref{fig:xrayfig} appear to be
dominated by soft (red) X-ray emission, a trend which is also apparent
in the count-rate statistics. For instance, all six are formally
detected in the soft band and four in the full band, but only one
source is detected in the hard band (source~2 in
Table~\ref{tab:sources} has $157\pm 52$ hard-band counts). Three
effects mitigate against detections in the hard band: (1) the higher
background rate in this band, (2) the smaller effective area in the
hard-band, and (3) the intrinsically soft nature expected for the
majority of extended X-ray sources. The band ratios (BRs) of the
sources, defined as the ratio of hard-band to soft-band count rates,
are listed in column~7 of Table~\ref{tab:sources} and are upper limits
in all but one case. The small BRs for the two brightest sources are
consistent with the low X-ray temperatures found for nearby X-ray
groups and poor X-ray clusters \citep[e.g.,][]{Xue2000}. The BRs for
the other four sources provide little constraint on potential spectral
models.

The two brightest X-ray sources have enough counts to provide more
detailed spectral constraints. X-ray spectra were extracted for each
source using the regions described in Table~\ref{tab:sources}. We used
only events taken at $-120\arcdeg$C for this spectral analysis since
$\approx$~75\% of the $\approx$~1~Ms survey was performed at this
temperature, and the particle background was $\approx$~20\% higher at
the time when $-110\arcdeg$C data were collected.\footnote{See
http://asc.harvard.edu/cal/Links/Acis/acis/Cal\_prods/bkgrnd/.} As
described in Paper~V, the CDF-N data have been corrected for radiation
damage using the procedure of \citet{Townsley2000}. To complement the
corrected data, we also used the modified response matrix files (RMFs)
and quantum efficiency uniformity files (QEUs) supplied with the
corrector. Since the physical positions of each extended X-ray source
on the ACIS-I CCDs varied among the 12 CDF-N observations, we
extracted individual RMFs and QEUs for each observation and averaged
them together weighting by the number of counts in each observation. 
The spectrum of each source was binned such that each spectral bin
contained at least 30 counts, and spectral fitting was performed using
{\sc xspec} \citep{Arnaud1996}. We only included energies where the
data are not dominated by the background, corresponding to
0.5--3.5~keV for source~2 and 0.5--5.0~keV for source~6. We fitted the
spectra with absorbed Mewe-Kaastra-Liedahl thermal plasma models
\citep[i.e., the {\sc xspec} {\sc mekal} model;][]{Mewe1985,
Kaastra1992, Liedahl1995} and obtained acceptable fits. Note that
Raymond-Smith thermal plasma models \citep{Raymond1977} were equally
acceptable. For both sources, the reduced $\chi^{2}$ values were
reasonable ($\chi^{2}_{\nu}=1.30$ for source~2 and 1.04 for source~6),
and no obvious systematic residuals were present. In the fits, the
plasma temperature was left as a free parameter, the absorption column
density was fixed at the Galactic value, and the redshift was fixed to
that given in Table~\ref{tab:sources}. No constraints could be placed
on the abundance values, so they were set to 0.3 times solar, a value
typical of nearby groups \citep[e.g.,][]{Mulchaey2000}. Since both of
these sources have low source-to-background ratios, we extracted
several different background regions of varying sizes and shapes to
determine what effect the background had on our spectral fits. 
Depending on the local background region used, we found that the
best-fit temperatures of the sources varied by at most 20\%, well
within the quoted errors. The results of the spectral fits are listed
in Table~\ref{tab:sources}.

Soft-band X-ray fluxes were determined using the best-fit model from
{\sc xspec}, or, in the case of the four faint sources, a $kT=1$~keV
thermal plasma model with an abundance of 0.3 times solar. The fluxes,
both for detections and upper limit estimates, were corrected for
vignetting and for the areas masked out to eliminate contaminating
point sources. The apparent angular sizes of all the objects lie
within the range 45--90$\arcsec$ (i.e., $\sim$100--600~kpc over the
expected redshift range), and only one has an extent
$\ga$~$60\arcsec$. Strictly speaking, the quoted extents of these
extended X-ray sources should be regarded as lower limits, since a
non-negligible fraction of their flux could lie in the outer isophotes
just below our detection threshold. This effect is most noticeable in
groups of galaxies, where the surface brightness profile is shallower
than that of clusters. For instance, isophotal measurements of X-ray
detected groups have been found to underestimate the true X-ray
luminosities of their hot gas by factors of up to 2--3 in some extreme
cases \citep[e.g.,][]{Helsdon2000}. Given the irregular morphologies
of the brighter sources and the limited statistics of the fainter
sources, we have not attempted to model their surface brightness
profiles nor have we performed any related luminosity corrections
(e.g., to the virial radius). X-ray luminosities were calculated
assuming the most likely redshifts of potential optical group or
cluster members (see $\S$\ref{optical}) and were corrected for
Galactic absorption.

\begin{figure*}
\vspace{8.5in}
\figcaption[fig2.eps]{{\it Chandra} ACIS-I adaptively smoothed 
soft-band contours of the six extended sources are overlaid on
$V,I,HK'$ color images (the $V,I,HK'$ images of Barger et al 1999
represent blue, green and red, respectively), with the exceptions of
sources~1 and 6, which lie outside the field of view of the color
image and instead are overlaid on the $R$-band image of
\citet{Liu1999}. Each contour indicates an increase in X-ray flux by a
factor of $\sqrt{2}$. The bar in a corner of each panel shows the
angular size on the sky of 100~kpc at the most likely redshift of the
extended source. Note that source~3 is likely to be associated with
the FR~I radio galaxy VLA~J1236440$+$621133, indicated by the second,
smaller set of VLA radio contours from
\citet{Richards1999a}. Abcissa denotes Right Ascension, ordinate denotes Declination.
\label{fig:overlay}} 
\end{figure*}

Based on the average properties of nearby clusters and groups of
galaxies \citep[e.g.,][]{Ponman1996, Mohr1997, Vikhlinin1998}, five of
the six extended CDF-N X-ray sources have apparent physical extents
(assuming the redshifts indicated in $\S$\ref{detections}) similar to
those of X-ray groups ($\sim$100--300~kpc). The properties of
source~2, however, appear to be more extreme; with an X-ray
temperature of at least 2~keV and a physical extent of $\ga$~600~kpc,
its X-ray properties are more consistent with those of a moderately
X-ray luminous cluster. Note that deviations from the assumed
redshifts above $z=0.3$ will not change the physical size by more than
a factor of two at most. Sources~2 and 6, the two brightest sources,
are both elongated and clumpy, suggesting either point-source
contamination or that they are unrelaxed systems perhaps undergoing
mergers.

We have thus far only entertained the possibility that these extended
X-ray sources are extragalactic, but could any of these sources be
Galactic in nature? Given that the angular sizes
(45$\arcsec$--90$\arcsec$) and, when measurable, temperatures
($\sim$2--4~keV) of the sources discussed here are quite different from
those of typical interstellar medium clouds
\citep[i.e., $\approx$~1$\arcdeg$ scales and $\approx$~0.25~keV
temperatures;][]{Snowden1998, Kuntz2000}, and furthermore that the
CDF-N is at high Galactic latitude ($b=54\fdg828$), this possibility
seems unlikely. Any other form of diffuse Galactic X-ray emission
(e.g., a supernova remnant) should have obvious extended optical or
radio counterparts.

\section{Optical Constraints}\label{optical}

\subsection{Optical Images and Photometry}\label{photometry}

To assess the optical nature of these six sources, we inspected all of
the publicly available images covering the CDF-N region. These include
the $HK^\prime$, $I$, $V$, and $B$-band images of
\citet{Barger1999}\footnote{These images are available at
http://www.ifa.hawaii.edu/$\sim$cowie/hdflank/hdflank.html.} and the
$U_{\rm n}$, $G_{\rm r}$, and ${\cal R}$-band images of
\citet{Steidel1993}, all of which cover an
$\approx9\arcmin\times9\arcmin$ region surrounding the HDF-N, as well
as the $R$-band image of \citet{Liu1999} which extends over the entire
CDF-N region. The $HK^\prime$, $I$, ${\cal R}$, $R$, $V$, $G_{\rm r}$,
$B$, and $U_{\rm n}$-band images have $\approx$~2$\sigma$ detection
limits of 21.2, 25.5, 25.6, 23.0, 26.5, 26.4, 26.3, and 25.0,
respectively. Figure~\ref{fig:overlay} shows contours of the
adaptively smoothed soft-band X-ray image for each of the six X-ray
sources overlaid on either the Barger et al. $I$-band image or, when
no $I$-band coverage was available, the wide-field Liu et al. $R$-band
image. The X-ray and optical astrometric reference frames have both
been tied to the FK5 radio coordinate grid using several dozen bright,
point-like radio sources; the typical resulting X-ray/optical offsets
are $\la0\farcs6$ for all point sources with off-axis angles less than
5$\arcmin$, and $\la1\farcs2$ for sources 5$\arcmin$--10$\arcmin$
off-axis.

Magnitudes for sources were determined using the {\sc SExtractor}
photometry package \citep{Bertin1996} with the ``Best'' magnitude
criteria, a $2\sigma$ detection threshold, and a 25-pixel Gaussian
wavelet. As a consistency check for the {\sc SExtractor} photometry,
we matched our sources to sources in the catalog of
\citet{Barger1999}; we found good agreement, with typical 1$\sigma$
magnitude deviations of $\la\pm0.25$. As expected, the largest
deviations were always near the detection threshold of the images.

\subsection{Tests of Clustering}\label{montecarlo}

In general, there appear to be several bright optical galaxies located
within the X-ray contours of each source, suggesting a moderate level
of clustering. Only source~2 fails to follow this trend. One of the
most effective tests for clustering is the ``red sequence'' method of
\citet{Gladders2000}, which relies on the assumption that {\it all}
significant real clusters have a red sequence of early-type galaxies
and that this sequence clearly stands out among the field galaxy
population at brighter magnitudes (e.g., $I\la22$--23) in $V-I$
colors. The best constraints using this method, however, rely on
precision photometry and morphological classifications, neither of
which are feasible with our current optical data. Nevertheless, we
used the $V$-band and $I$-band images which cover
$\approx9\arcmin\times9\arcmin$ to generate $V-I$ versus $I$
color-magnitude diagrams identical to those presented in
\citet{Gladders2000}. We found no indication of a red sequence 
in any of the four extended CDF-N sources that lie within these
images, nor for ClG~1236$+$6215, an optically-selected cluster. Our
results suggest that either better photometry and morphological
classifications are required or that the universality of the red
sequence in clusters does not necessarily extend down to the level of
groups or poor clusters.

While less efficient, another method to test for optical overdensities
near the extended X-ray sources is Monte Carlo simulations on
the $I$-band and $R$-band images. In these simulations an aperture was
positioned at random on the image and the number of galaxies above a
given magnitude within it was tallied. For each of the extended X-ray
sources, we adopted a circular aperture equal to twice the size of the
major axis listed in column~3 of Table~\ref{tab:sources}; since the
detectable X-ray emission from groups of galaxies often only extends
out to a fraction of the virial radius, many potential optical group
members may lie outside of the region traced by the X-ray emission
\citep{Mulchaey2000}. For the $I$-band image, the magnitude limit was
set to $I=24.0$. For the $R$-band image, the magnitude limit was set
slightly above the detection threshold to $R=22.5$. Unfortunately, the
combination of the large apertures used and the relatively small
angular sizes of the $I$-band and $R$-band images
($\approx9\arcmin\times9\arcmin$ and
$\approx30\arcmin\times30\arcmin$, respectively) severely limit the
number of statistically independent cells we were able to use for
Monte Carlo simulations; for the $I$-band and $R$-band images we used
$\sim$80 and $\sim$900 trials per CDF-N extended source, respectively.

Column~11 of Table~\ref{tab:sources} gives the results of the Monte
Carlo simulations, listed as the fraction of trials containing fewer
galaxies than found in an aperture at the position of the X-ray source
($P_{\rm over}$). Simulations using the $I$-band image yielded
significant ($>$~90\% confidence level) overdensities for sources~3,
4, and 5, although the small number of independent cells restricts the
precision of these simulations. For comparison, we also ran
simulations on a region near the HDF-N known to have a large
overdensity of sources, the optical cluster of galaxies
ClG~1236$+$6215 (see $\S$\ref{intro} and $\S$\ref{nondetections}). We
found that the number of sources in ClG~1236$+$6215 within a
30$\arcsec$ radius aperture, when compared to the 80 Monte Carlo
simulation trials, was overdense at the 100\% confidence level. Since
the $I$-band simulations do not cover the regions around sources~1 and
6, simulations using the $R$-band image were performed instead. The
$R$-band simulations yielded strong results for both sources and
confirmed our $I$-band results. We note that decreasing the aperture
radius used in the simulations to that of the X-ray extraction major
axis yielded similar results, while varying the magnitude criterion by
one magnitude brighter or fainter yielded mixed results. A more
detailed study of the surface density of optical sources both as a
function of magnitude and radius may improve the statistics but is
beyond the scope of this work. These simulation results imply possible
optical clustering for five of the six sources. Since distant groups
are generally difficult to distinguish from background sources based
on statistical deviations alone, our null results for source~2 may not
necessarily be inconsistent with its identifications as a potential
cluster.

\subsection{Redshift Information}\label{redshifts}

Clusters and groups of galaxies are usually identified through the
redshifts of their optical members. We estimated the redshifts of the
extended X-ray sources using the photometric redshift catalog of
Fern{\' a}ndez-Soto, Lanzetta, \& Yahil (1999) and the spectroscopic
redshift catalogs of \citet{Cohen2000}, \citet{Dawson2001}, and A.~E. 
Hornschemeier et al., in preparation. Sources~3 and 6 have clear
redshift identifications (see $\S$\ref{detections}). Unfortunately,
all of the extended X-ray sources apart from source~3 lie outside the
well-studied region of Cohen et al. (i.e., $>$~700 redshifts within a
$\sim$$4\arcmin$ radius region centered on the HDF-N), so
spectroscopic coverage was often limited to only a few sources in each
field. To augment these published redshifts, we estimated photometric
redshifts for all optical sources that have extensive multi-band
optical coverage using the optical images noted in
$\S$\ref{photometry}. Specifically, we found that our photometric
redshift estimates were most reliable for sources with detections in
the $HK^\prime$-band (i.e., $HK^\prime\la21.2$) and in at least four
of the other six optical bands (the wide-field $R$-band image was not
included since it provided no new information). There were 858 sources
which satisfied this constraint. Our requirement of a source detection
in the $HK^\prime$-band, rather than some other band, was based on the
fact that a $HK^\prime$ detection provided much stronger redshift
constraints from the spectral template fitting procedure compared to
any of the other bands and resulted in a lower percentage of
catastrophic failures (see below). By necessity, all 858 of these
sources lie in the $\approx9\arcmin\times9\arcmin$ area centered on
the HDF-N.

To measure photometric redshifts we used the publicly available
photometric redshift code {\sc hyperz} Version 1.1
\citep{Bolzonella2000}.\footnote{See
http://webast.ast.obs-mip.fr/hyperz/.} In performing the photometric
redshift fitting, we used both the \citet{Coleman1980} and
\citet{Bruzual1993} spectral templates provided with {\sc hyperz} and
allowed up to 1 mag of visual extinction. Of the 858 source redshifts
estimated with {\sc hyperz}, 455 have published spectroscopic
redshifts. Considering that the ensemble of images were obtained with
different telescopes, instruments and observing conditions, and,
furthermore, that the optical bands were not optimized for estimating
photometric redshifts, we find notably good agreement between the
photometric and spectroscopic measures. Only about 5\% of the sources
fail catastrophically (i.e., $|z_{\rm sp}-z_{\rm ph}|>1.0$); the
majority of these sources are photometrically estimated to lie at high
redshifts ($z_{\rm ph}|>1.5$), but are in fact nearby ($z_{\rm
sp}<0.5$) galaxies with $U_{n}-HK^\prime<3.0$ and are usually
classified by \citet{Cohen2000} as composite spectral types with both
emission and absorption features typical of dwarf starburst galaxies. 
These sources appeared to fail because {\sc hyperz} did not have a
suitable spectral template to use in the model fitting. If we exclude
sources with $U_{n}-HK^\prime<3.0$ and $|z_{\rm ph}|>1.5$ (59 in all,
28 with spectroscopic redshifts), we find a 1$\sigma$ deviation
between the spectroscopic and our photometric redshifts of $\Delta z =
0.15$.\footnote{A catalog of these 799 sources is provided at
http://www.astro.psu.edu/users/niel/hdf/hdf-chandra.html.} 
Figure~\ref{fig:photoz} shows a comparison of the photometric and
spectroscopic redshifts, both as a scatter plot of individual sources
and as a histogram of the difference between the two measures binned
in 0.1 redshift deviation intervals. Only sources with photometric or
spectroscopic redshifts below 1.5 are plotted in since the spectral
constraints imposed by the detection limits of the images and the
requirement for a $HK^\prime$-band detection naturally exclude most
high-redshift sources.

The photometric redshift estimates described above have allowed us to
enhance the redshift coverage in half of the CDF-N source fields (2,
4, and 5) and have led to an improved distance constraint for source~2
(see $\S$\ref{detections}). Even with these additional redshift
estimates, we are only 
confident in the redshift determinations for sources~3 and 6. To give
the reader a feeling for how we made our redshift estimates, we show
in Figure~\ref{fig:redshift} optical images for the two likely high
redshift sources (2 and 3) with spectroscopic and photometric
\citep[ours and][]{Fernandez1999} redshifts overlaid. For source~2,
which could lie at one of several possible redshifts, we plot
spectroscopic and photometric redshifts of individual galaxies when
they are consistent to within their measurement error with three
possible redshifts of the X-ray source at $z=0.68$, 0.80, and 1.35
(note that 14 out of a total 36 redshifts have been plotted). For
source~3, we have only plotted redshifts of individual galaxies when
they are consistent to within their measurement error with the likely
redshift of the X-ray source at $z=1.01$ (note that 32 out of a total
77 redshifts have been plotted). For this, we assumed that the
spectroscopic redshifts have errors of $\Delta z_{\rm sp}=0.01$, that
the photometric redshifts of
\citet{Fernandez1999} have errors of $\Delta z_{\rm ph}=0.09$, and
that the photometric redshifts determined here have errors of $\Delta
z_{\rm ph}=0.15$. These images are described in more detail in
$\S$\ref{detections}.

\vspace{0.2cm} 
\vbox{
\centerline{
\hglue-2.8cm{\includegraphics[width=11.5cm]{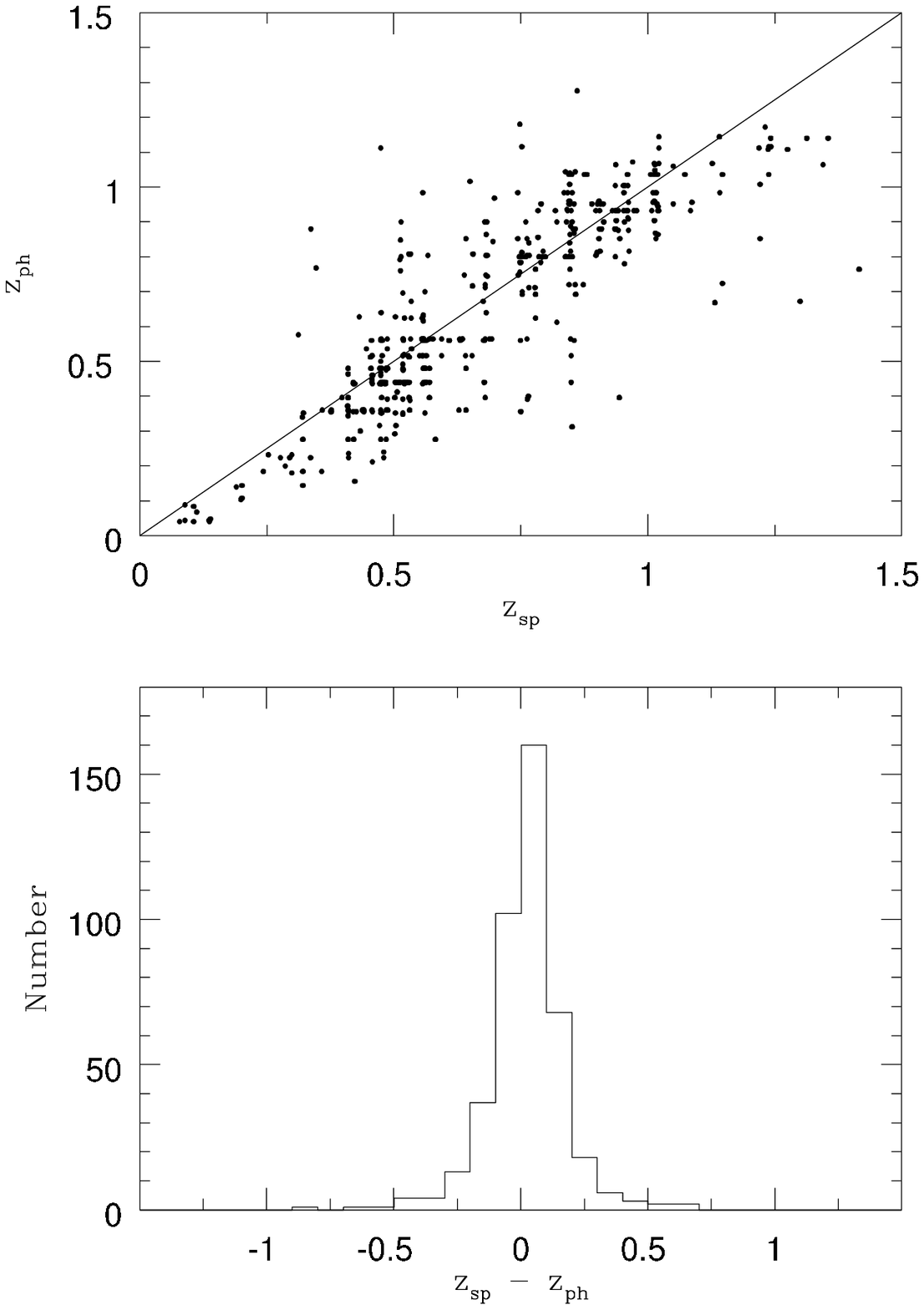}}
}
\vspace{-4.0cm} 
\figcaption[fig3.eps]{A comparison of our photometric redshift estimation 
($z_{\rm\ ph}$) for 427 sources in the HDF-N and environs which also
have spectroscopic ($z_{\rm\ sp}$) redshifts of \citet{Cohen2000},
\citet{Dawson2001}, and A.~E. Hornschemeier et al., in preparation. 
The upper panel shows a comparison of individual sources, and the
lower panel shows a histogram of the difference between the two
redshift measures binned in 0.1 intervals. Not plotted are the 28 (out
of 59 total) catastrophic failures defined as sources with
$U_{n}-HK^\prime<3.0$ and $z_{\rm ph}|>1.5$, which presumably fail due
to the limited number of spectral templates used (see
$\S$\ref{redshifts}). The 1$\sigma$ deviation of the photometric
redshift from the spectroscopic redshift is $\Delta z = 0.15$.
\label{fig:photoz}}
}

\begin{figure*}
\vspace{7.75cm}
\vspace{-0.5cm}
\figcaption[fig4.eps]{Sources with spectroscopic or photometric
redshifts meeting the criteria given in $\S$\ref{redshifts} are overlaid
on the optical color images of sources~2 and 3 also shown in
Figure~\ref{fig:overlay}. The dashed lines indicate the outer X-ray
contour shown in Figure~\ref{fig:overlay}. Sources with photometric
redshifts are denoted with an appended ``p''. The pairs of lines in
the image of source~2 indicate the positions of the six unusual
objects described in $\S$\ref{detections}. Source~3 is likely to be
associated with the $z=1.013$ FR~I radio galaxy VLA~J1236440$+$621133
shown by the arrow in that panel. All of the photometric redshifts in
the upper half of the source~3 image come from the HDF-N estimates of
\citet{Fernandez1999}, since the HDF-N partially overlaps with this
region; some of these sources are detected only in the HDF-N imaging
and appear blank in this image. Note that there are five sources
concentrated near VLA~J1236440$+$621133 with $z_{\rm\hbox{   \tiny ph}}=0.960$; to
avoid visual clutter, these sources are indicated by squares.
Abcissa denotes Right Ascension, ordinate denotes Declination.
\label{fig:redshift}}
\end{figure*}

\section{Individual Source Notes}\label{notes}

This section is devoted to descriptions of the multi-wavelength
properties of each source. We first discuss the two clear detections,
sources~2 and 6, and then the other four detected sources for which
less diagnostic information is available. Lastly, we provide
constraints on two undetected sources for which X-ray emission was
expected based on observations at other wavelengths.

\subsection{Detected X-ray Sources}\label{detections}

{\it Source~2 --- CXOHDFN~J123620.0$+$621554}: This source has a
double-peaked X-ray morphology suggestive of an ongoing merger and a
best-fit plasma temperature of $kT\approx3.7$~keV. Unlike the other
extended X-ray sources in the CDF-N, source~2 has no obvious optical
counterparts in the $I$-band image, suggesting a fairly high redshift. 
There are a number of nearby sources with spectroscopic or photometric
redshifts consistent with $z=0.68$ or 0.80. At such redshifts, this
object would have unabsorbed soft-band luminosities of
$3.2\times10^{42}$\xlum\ or $4.8\times10^{42}$\xlum, respectively, and
it would likely be an underluminous X-ray cluster or a moderately
luminous group of galaxies. Very few of the sources with redshifts,
however, lie within the innermost X-ray contours (see
Figure~\ref{fig:redshift}). Focusing only on the optical sources that
{\it do} lie close to the center of the brighter X-ray peak, we find
that 123620.1$+$621555, an $I=22.5$ source at $z=1.35$, lies
$\approx5\arcsec$ from the X-ray peak. This source has a narrow
[\ion{O}{2}] $\lambda$3727 emission line and \ion{Mg}{2} absorption
features \citep{Dawson2001}; its broad-band colors and absolute
magnitude suggest that it may be atypically bright, possibly due to
AGN activity. We also identify 123620.1$+$621555, an $I=24.0$ galaxy,
to be coincident with the X-ray peak. Either source could plausibly be
a candidate ``Brightest Cluster Galaxy'' (BCG) within the cluster.
BCGs are thought to be excellent distance indicators
\citep[e.g.,][]{Postman1995}; assuming 123620.1$+$621555 is a typical
BCG with $M_{\rm R}$ between $-22.0$ and $-23.0$ based on the
compilation of \citet{Hoessel1985}, and using the E1 galaxy
$K$-correction model and $R-I$ colors of
\citet{Poggianti1997}, we find that this source should lie at a
redshift of $z=1.3$--1.5.\footnote{Note that because of the strong
$K$-correction for elliptical galaxies at these high redshifts, a BCG
may not in fact be the galaxy with the brightest apparent magnitude in
the cluster.} The proximity of these two sources to the X-ray peak
favors a high redshift for the X-ray source.

Another notable characteristic of this extended X-ray source is the
high density of unusual objects found clustered within a $45\arcsec$
radius of its brightest X-ray peak (see Figure~\ref{fig:redshift}). 
Specifically, there are two Very Red Objects (VROs) defined as
$I-K\ge4.0$ \citep{Alexander2002} and four optically faint radio
and/or X-ray sources with $I\ge24$ \citep{Richards1999b,
Alexander2001}. From the combined source densities of VROs, optically
faint radio sources, and optically faint X-ray sources, we would
expect only 0.8 such sources. Interestingly, four of these six unusual
objects lie within a $15\arcsec$ radius of the X-ray peak, suggesting
an association. Furthermore, one of the optically faint radio sources
has a submillimeter detection and a millimetric redshift of $z_{\rm
mm}=1.8^{+0.7}_{-0.6}$ \citep{Barger2000}. This redshift is consistent
with those of the two optical galaxies near the X-ray peak. The rest
of the objects are expected to lie at redshifts of $z\sim1$--3.

Overdensities of this kind have been noted for several other
high-redshift clusters \citep[e.g.,][]{Chapman2000, Smith2001}; two
plausible scenarios can explain such an overdensity. The first is that
source~2 is physically associated with the VROs and optically faint
radio and X-ray sources, and that they all lie at a redshift of
$z\sim1$--3. At such redshifts, the rest-frame unabsorbed soft-band
luminosity of this source would be
$\approx(0.8$--$12)\times10^{43}$\xlum. An alternative scenario is
that the X-ray source lies at an intermediate redshift, as suggested
by the spectroscopic and photometric redshifts of nearby sources, and
the overdensity of unusual objects is a manifestation of gravitational
lensing. We note that a similar extended X-ray source was found by
\citet{Hasinger1998} within the Lockman Hole: the double-peaked X-ray
source, RX~J105343$+$5735, thought to lie at $z=1.267$.
RX~J105343$+$5735 has both an overdensity of very red ($R-K^\prime$ =
5.5--5.7) objects nearby and exhibits evidence for gravitational
lensing in the form of a bright arc near the center of the X-ray peak
\citep{Lehmann2000, Lehmann2001}. The only differences between
RX~J105343$+$5735 and source~2 are that the former has clear optical
counterparts at $R\sim21$--$22$ (nearly 3--4 magnitudes brighter than
found for source~2) and the latter exhibits no obvious signs of
lensing. Further observations of this object with {\it Chandra} and
{\it HST} should help to reveal its nature. At the aforementioned
redshifts, the physical size of source~2 would be $\ga$~600~kpc.

{\it Source~6 --- CXOHDFN~J123756.0$+$621506}: Like source~2, this
source is also clearly elongated and has two distinct peaks of
emission, suggesting a possible merger. The brighter of the two X-ray
peaks coincides with three optically bright, nearly overlapping
galaxies in the $R$-band image of \citet{Liu1999}. The optical
magnitudes of these galaxies imply that the extended X-ray source is
likely to be a typical X-ray emitting group at moderate redshift. One
interesting aspect of source~6 is that the second X-ray peak shown in
Figure~\ref{fig:overlay} lacks a bright optical counterpart down to
the $R$-band limit. Such a morphology has few analogs among nearby
X-ray groups (e.g., Mulchaey et al. 1996; J. Mulchaey 2001, private
communication) and suggests that there may be contamination from one
or more faint X-ray point sources. From inspection of the unsmoothed
X-ray image, there appear to be no obvious point sources. However,
given that source~6 lies at an off-axis angle of $\sim$8$\arcmin$,
X-ray photons from any faint point sources would be distributed over a
diameter of $\approx$~10$\arcsec$ (i.e., the 1.5~keV 50\% encircled
energy radius) and would not necessarily be detectable by eye. 
Therefore, to search for possible contamination from faint X-ray point
sources, we adaptively smoothed the raw X-ray images, but restricted
the smoothing algorithm to smooth only on scales smaller than the {\it
Chandra} PSF (i.e., 5$\arcsec$). A comparison of this minimally
smoothed X-ray image and the $R$-band image confirms that the second
X-ray peak lies in between two $R\sim$22--23 magnitude galaxies
(123756.5$+$621456 and 123758.8$+$621458), both with hints of X-ray
emission. Thus this second peak is likely to be an artifact of the
adaptive smoothing algorithm. Moreover, these galaxies are likely to
be related to the X-ray group, since the probability that two unrelated
$R\sim$22--23 magnitude X-ray sources (with estimated fluxes of
$\sim$1$\times10^{-16}$\xflux) would be found within a 15$\arcsec$
radius of this X-ray group is less than 1\%.

Since source~6 lies $\sim$$8\arcmin$ from the center of the
HDF-N, there are no published spectroscopic redshifts for any of the
galaxies in this field. Therefore, to determine its distance, moderate
resolution long-slit spectroscopic observations were made with the
Hobby-Eberly Telescope (HET) Marcario Low Resolution Spectrograph
\citep[LRS;][]{Hill1998, Hill2000} for two of the three optical
galaxies which are most likely to be associated with the X-ray source
(see Figure~\ref{fig:overlay}). The results of these observations are
presented in Appendix~\ref{app:HET} and show that both galaxies lie at
$z=0.190$. Thus the extended emission from this source has a physical
size of $\approx$~150~kpc and a rest-frame unabsorbed soft-band
luminosity of $2.1\times10^{41}$\xlum, properties consistent with poor
groups of galaxies studied locally \citep[e.g.,][]{Mulchaey1996,
Ponman1996, Helsdon2000}. At this redshift, it is one of the more
distant X-ray groups known. From our spectral analysis in
$\S$\ref{properties}, source~6 has a derived a rest-frame plasma
temperature of $kT\approx2.9$~keV. This is much higher than expected
for a group of this size and luminosity and suggests the possibility
of either point-source contamination, shock-heating from infalling
gas, or non-gravitational heating. Considering the detection of two
faint galaxies near the second X-ray peak, point source contamination
appears to be the most likely explanation.

{\it Other Sources}: Sources~1, 3, 4, and 5 have less well defined
X-ray properties. All appear to have roughly spherical X-ray
morphologies, although spatial irregularities cannot be ruled out
given their low source-to-background ratios. Moreover, only weak
constraints can be placed on their X-ray spectral nature. Source~3 is
particularly notable because it lies within the HDF-N itself near the
$z=1.01$ FR~I radio galaxy VLA~J123644.3$+$621133 and several other
$z\approx 1.01$ objects \citep{Richards1998, Fernandez1999,
Cohen2000}. In Paper~IV, the FR~I itself was detected (as
CXOHDFN~J123644.3$+$621132), and at the time no nearby extended X-ray
emission was found, despite the fact that FR~I radio sources are often
associated with clusters of galaxies. Now with the entire 1~Ms
dataset, it appears that such emission was present just below the
prior detection threshold. The FR~I lies within the extent of the
extended emission but appears somewhat offset from its poorly defined
center. \citet{Cohen2000} found 23 objects clustered around
$z\approx1.01$ in a $\sim4\arcmin$ radius, implying a possible
``filament'' structure. In Figure~\ref{fig:redshift}, we show 32
galaxies with spectroscopic and photometric redshifts within the
vicinity of source~3. At $z=1.01$, the extended source has an
unabsorbed soft-band luminosity of $2.0\times10^{42}$\xlum\ and an
apparent physical size of about 200~kpc (see
Figure~\ref{fig:overlay}). Thus it is likely to be a moderately
luminous X-ray group or possibly a poor X-ray cluster.

Less is known about the other three sources, and none has
a clear redshift. Based on the limited spectroscopic and photometric
redshift information available and the magnitudes of the bright
sources coincident with the X-ray emission, sources~1 and 4 may lie at
$z\approx0.44$ and $z\approx0.38$, respectively, while source~5 is
thought to lie at $z\sim0.4$--0.7. At such distances, these sources
would have unabsorbed soft-band luminosities of
$\approx$(2--9)$\times10^{41}$\xlum\ and apparent physical sizes of
$\approx$~200--300~kpc. As more X-ray and optical data become
available, the nature of these sources should be revealed.

\subsection{Cluster Candidates not Detected in the X-ray Band}\label{nondetections}

{\it ClG~1236$+$6215}: This optical cluster was originally noted by
\citet{Barger1999} as an overdensity of red objects, but it recently has been 
verified as a cluster by \citet{Dawson2001} based on
spectroscopic follow-up of several bright sources within $45\arcsec$
of $\alpha=12^{\mathrm h}36^{\mathrm m}39\farcs6$, $\delta=
+62\arcdeg15\arcmin54\arcsec$. The cluster lies at a redshift of
$z=0.85$. Based on the line-of-sight velocity dispersion calculated by
\citet{Dawson2001} and the X-ray luminosity-velocity dispersion
relation of \citet{Xue2000}, Dawson et al. calculate that the
bolometric X-ray luminosity of ClG~1236$+$6215 should be
$L_{\rm X}\approx1.2\times10^{44}$\xlum, or $\approx$~$3\times10^{43}$\xlum\ in
the 0.5--2.0~keV band assuming a 6~keV thermal plasma model
\citep[e.g.,][]{Xue2000}. Unfortunately, 
this cluster lies quite close to a CCD chip gap, so depending upon its
exact shape and extent, portions of it may only be exposed for about
half of the total exposure time. While {\sc vtpdetect} did not
formally detect the cluster, there are hints of diffuse X-ray emission
in the adaptively smoothed soft-band image. 
Figure~\ref{fig:clusterover} shows the $I$-band image of
ClG~1236$+$6215 with the X-ray contours overlaid. Based on its X-ray
appearance, we extracted counts for this putative source using a
circular aperture with a $30\arcsec$ radius, taking care to exclude
the point source CXOHDFN~J123642+621546 detected in Paper~V. No
significant detection was found, resulting in the 3$\sigma$ upper
limits listed in Table~\ref{tab:nosources}. At a redshift of $z=0.85$,
the limiting soft-band and full-band luminosities are
$2.1\times10^{42}$\xlum\ and $3.9\times10^{42}$\xlum, respectively. 
The soft-band X-ray luminosity limit is at least a factor of 15 less
than predicted by Dawson et al. and, if this source {\it does} have
X-ray emission, it would have to be a very poor X-ray emitting cluster
of galaxies \citep[e.g., compare with the sample of][]{Burns1996}.

{\it VLA~J123725.7$+$621128}: This is one of only two VLA-detected
sources in the vicinity of the HDF-N which shows extended radio
emission on extragalactic scales \citep{Richards1998}; the other is
VLA~J123644.3$+$621133 associated with source~3 described in
$\S$\ref{detections}. With a 1.4~GHz flux density of 6~mJy, this
object is one of the brightest radio sources in the field and is
classified morphologically as a WAT source. It is optically identified
with an $I=22.9$ elliptical galaxy and is estimated by
\citet{Hornschemeier2001} and \citet{Snellen2001} to lie at a redshift of $z\sim1$--2
based on the the $K$-$z$ relation. WAT sources are often associated with
central dominant ellipticals in rich clusters of galaxies
\citep[e.g.,][]{Rudnick1976, Burns1994, Gomez1997}, so it is surprising
there are no hints of extended X-ray emission within its vicinity. 
Table~\ref{tab:nosources} lists the 3$\sigma$ upper limit derived
using a $30\arcsec$ radius aperture centered on the radio source. At
$z\sim1$--2, the limiting soft-band and full-band luminosities are
$(2.6$--$14.3)\times10^{42}$\xlum\ and $(5.2$--$28.1)\times10^{42}$\xlum,
favoring comparisons with X-ray weak clusters or typical X-ray groups.

\vspace{0.6cm}

\centerline{
\vspace{7.3cm}
}
\vspace{-0.4cm}
\figcaption[fig5.eps]{{\it Chandra} ACIS-I adaptively 
smoothed soft-band contours of the undetected cluster of galaxies ClG
1236$+$6215 overlaid on a $V,I,HK'$ color image (the $V,I,HK'$
images of Barger et al 1999 represent blue, green and red,
respectively). The properties of the image (e.g., contours and bar)
are identical to those described in Figure~\ref{fig:overlay}. There is
apparent X-ray emission but no formal detection (see
$\S$\ref{nondetections}). Abcissa denotes Right Ascension, ordinate
denotes Declination.
\label{fig:clusterover}} 

\section{Discussion and Conclusions}\label{discuss}

\subsection{Basic Nature of the Extended X-ray Sources}\label{basic}

The general X-ray and optical characteristics of the extended sources
in the CDF-N (i.e., their soft X-ray luminosities, apparent X-ray
sizes, and weak optical clustering) are most comparable to those of
nearby groups of galaxies \citep[e.g.,][]{Mulchaey1996, Ponman1996,
Helsdon2000}. The only exception appears to be source~2, which has a
larger angular size and contains an overdensity of unusual objects; it
is likely to be a poor-to-moderate X-ray cluster at high redshift
(i.e., $z\ga0.7$). This CDF-N observation has also allowed us to place
strong constraints on two potential extended X-ray emitting systems,
ClG 1236$+$6215 and VLA~J123725.7$+$621128. Both systems lie at high
redshift and, from their undetected status, may still be in the early
stages of dynamical evolution, having not yet formed a central
concentration of hot gas and dark matter massive enough to produce
detectable X-ray emission \citep[e.g.,][]{Blanton2001, Donahue2001}.

For the two sources with enough counts to perform detailed spectral
analysis (sources~2 and 6), we have modeled their X-ray spectra and
estimated thermal plasma temperatures. Comparing the most likely
rest-frame X-ray luminosities and temperatures of these two sources with
the well-established $L_{\rm X}-kT$ relation for clusters and groups
\citep[e.g.,][]{Allen1998, Xue2000}, we find that the X-ray
temperature of source~2 is consistent with its bolometric luminosity
if the X-ray source lies at high redshift (i.e., $z\ga0.7$), while the
X-ray temperature of source~6 is clearly too high to be consistent
with its bolometric X-ray luminosity \citep[compare to Figure~1
of][]{Xue2000}. Whether the high temperature of source~6 is due
point-source contamination or non-gravitational heating mechanisms
such as star formation from individual group members or shock heating
of the infalling gas \citep[e.g.,][]{Metzler1994, Ponman1996,
Cavaliere1997} cannot be resolved with our current X-ray data.

Some additional knowledge about the state of the X-ray emitting gas in
source~2 can be gained from its observed X-ray morphology. The source
appears to be irregular and double-peaked, suggesting a young merger. 
Unlike source~6, this double-peaked source appears to retain its
extended, bimodal structure even in the minimally smoothed X-ray
images (e.g., similar to the images made for source~6 in
$\S$\ref{detections}) and is likely to be real.\footnote{Although we
cannot conclusively exclude that possibility of some faint,
point-source contamination.} Recent simulations of offset merging
clusters \citep{Ricker2001} suggest that during the evolution of
merging clusters, there is a short-lived phase of increased luminosity
and temperature. Furthermore, for large impact parameters, the
morphological structure of the merger remnant becomes bimodal. While
this phase is likely to last at most only a few Gyrs, the strong
enhancement in luminosity may help to offset their low observational
occurrence. Perhaps the bimodal structure we see in this source~2, as
well as in that of the Lockman Hole source RX~J105343$+$5735, are a
natural consequence of this presumably common, albeit short-lived,
stage in the formation of clusters and groups.

\subsection{Number Density of Extended X-ray Sources}\label{density}

The six detected sources are all found in the ``high-exposure'' area
($\approx$~130 arcmin$^{2}$; see $\S$\ref{data}) with exposure times
above 800~ks, implying an extended-source surface density of
167$^{+97}_{-67}$ deg$^{-2}$ (1$\sigma$) at a limiting soft-band flux
of $\approx3\times10^{-16}$\xflux. 
\vbox{
\vspace{-1.65cm} 
\centerline{
\hglue0.1cm{\includegraphics[width=9cm]{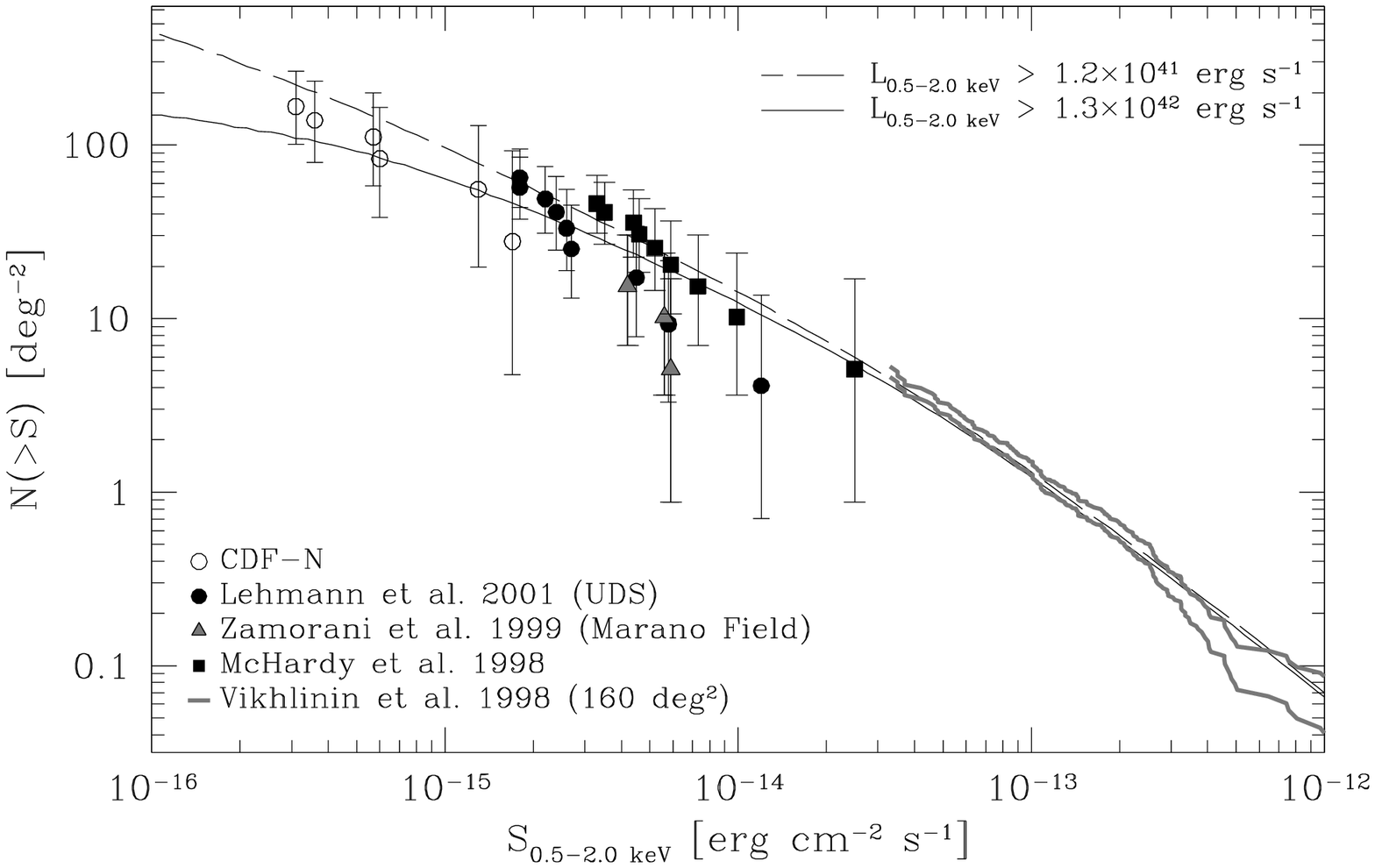}}
}
\vspace{-1.5cm} 
\figcaption[fig6.eps]{Cumulative number counts as a function of 
soft-band X-ray flux. Plotted are the extended sources from this
sample as well as those of \citet{McHardy1998}, \citet{Vikhlinin1998},
\citet{Zamorani1999}, and \citet{Lehmann2001}. The solid and dashed
lines show two models of the number counts derived from integrating
the XLF of \citet{Ebeling1997} above $L_{\rm
0.5-2.0~keV}=1.3\times10^{42}$\xlum\ and $1.2\times10^{41}$\xlum,
respectively (see $\S$\ref{density} for
details).\label{fig:logN-logS}}
}
\vspace{0.4cm} 

\noindent Figure~\ref{fig:logN-logS} shows
the cumulative soft-band number counts for \hbox{CDF-N} extended
sources (open circles). Also shown are the number counts from the {\it
ROSAT} extended-source samples of \citet{McHardy1998},
\citet{Vikhlinin1998}, \citet{Zamorani1999}, and \citet{Lehmann2001}. 

For comparison, we calculated the expected number density of X-ray
clusters using the Schechter expression from the bolometric
parameterization of the local cluster X-ray luminosity function (XLF)
by \citet[][see their Table~1]{Ebeling1997}. Two calculations were
made. In both cases, we assumed no evolution in the $\Omega_{\rm
M}=\onethird$, $\Omega_{\Lambda}=\twothirds$ cosmology, and the
integration was made over the range $z=0.015$--1.2. We assigned a
plasma temperature to each cluster assuming its bolometric luminosity
follows the $L_{\rm X}-kT$ relations of \citet{Xue2000}. The solid
line in Figure~\ref{fig:logN-logS} denotes the expected number counts
for sources with $L_{\rm X}>2.5\times10^{42}$\xlum\ (bolometric, or
equivalently $1.3\times10^{42}$\xlum\ in the 0.5--2.0~keV band for
$kT=1$~keV). Above this luminosity, the Ebeling et al. XLF is
supported by observational data \citep[see also][]{Rosati1998}. Only
two of the six CDF-N extended sources (or 55 sources deg$^{-2}$) are
likely to lie above this luminosity limit; this number density is
consistent with the predicted model. The dashed line in
Figure~\ref{fig:logN-logS} shows an extrapolation of the Ebeling et
al. XLF to include objects with $L_{\rm X}>2.0\times10^{42}$\xlum\
(bolometric, or $1.2\times10^{41}$\xlum\ in the 0.5--2.0~keV band for
$kT=0.5$~keV). Note that systematic studies of extended X-ray sources
with $L_{\rm X}\la5.0\times10^{42}$\xlum\ have been hindered by their
faint X-ray fluxes even locally, so this XLF extrapolation, while
plausible, is somewhat uncertain. Even so, these faint sources appear
to be consistent with the predicted model. Thus the CDF-N sources
appear to be consistent to within error with no evolution in the XLF. 
We caution, however, that our results may be affected by ``cosmic
variance'' due to the small field of view of this observation. 
Observations of larger-area fields with {\it Chandra} and {\it
XMM-Newton} are needed to confirm this result. 

Most of the extended sources in the CDF-N are likely to be groups,
and, as such, are thought to be affected more by energy and momentum
feedback from the stellar winds and supernovae following star
formation than by cosmological parameters
\citep[e.g.,][]{Cavaliere2000}. In fact, such pre-heating phase is
currently the best method for reconciling the different $L_{\rm X}-kT$
relations of groups and clusters. The lack of evolution in the
extended source number counts suggests that this pre-heating phase
may occur at redshifts higher than we tentatively observe here.

These findings also show that extended sources with fluxes
$\la10^{-15}$\xflux\ are not likely to contribute significantly to the
0.5-2.0~keV cosmic X-ray background \citep[$\la0.7$\% using the
results of][]{Cowie2002}. By comparison, extended sources with fluxes
$\ga10^{-14}$\xflux\ are thought to contribute $\sim$10\% to the
cosmic X-ray background \citep[e.g.,][]{Oukbir1997, Rosati1998}. The
fact that we do not detect extremely luminous sources is
understandable given their low surface density and that the HDF-N was
chosen to avoid such objects.

\subsection{Coincidence of X-ray Emission with Other Clustering Indicators}\label{peaks}

X-ray emission is thought to be an efficient method for tracing
intermediate-density structure within the ``cosmic web'' of
large-scale structure \citep[e.g.,][]{Tully1987, Bond1996, Pildis1996,
Cen1999, Pierre2000}. The extensive multi-wavelength observations
within the HDF-N and surrounding regions afford us an opportunity to
compare the clustering seen at X-ray wavelengths with that seen using
optical-to-near-IR and radio methods. We find significant
overdensities of optical galaxies around most of the CDF-N extended
X-ray sources from Monte Carlo simulations, but see little evidence
for any strong red sequence of early type galaxies and, perhaps
surprisingly, a stark contrast between the X-ray and optical
appearances of detected clusters in Figure~\ref{fig:overlay}. Unlike
most local clusters or groups, very few of the extended X-ray sources
appear to be associated with obvious optical clusters or groups. This
outcome is quite understandable, given the dramatic increase in the
field galaxy number density at these magnitudes, but highlights a
potential deficiency in relying on optical methods alone.

Considering the two sources with secure redshifts, source~3 falls on
one of the redshift peaks noted by \citet{Cohen2000} and
\citet{Dawson2001}, and source~6 does not. This suggests that at least
some low-level clustering occurs within these observed ``filaments''
of galaxies. The redshifts of the four other extended X-ray sources
are still unknown and may possibly coincide with known overdensities. 
If the coincidence of X-ray groups and clusters with known
``filaments'' in the CDF-N region is anything like that in the {\it
ROSAT} North Ecliptic Pole Survey, we should expect at least 25\% (or
$\sim$2) of the extended X-ray sources to lie within such large-scale
structures \citep[e.g.,][]{Burg1992, Brinkmann1999, Mullis2001}. 
Alternatively, we find no evidence for X-ray emission associated with
the optically detected cluster ClG 1236$+$6215.

At radio wavelengths, two FR~I radio sources are known
within the CDF-N region. These sources are predominantly thought to
reside in or near rich clusters of galaxies. We find that one of these
is radio objects is coincident with an extended X-ray source near the
threshold of our current {\it Chandra} observation, while the other
remains undetected (although X-ray and optical constraints cannot rule
out a $z>1$ cluster).

Clearly the spatial coverage of the CDF-N is too small to permit firm
conclusions about selection biases between these different cluster
indicators. However, the analysis presented here does suggest that
relying on any single technique may fail to detect some (and possibly
many) potential clusters and groups.

\subsection{Future Work}\label{future}

We have provided here an extremely deep view of the faint extended
X-ray source population in the CDF-N. Several details about these
objects are revealed, particularly for the brightest two sources, one
of which appears to be bimodal and is perhaps undergoing a merger. 
While the six CDF-N objects generally confirm our understanding of
less distant clusters and groups, they also provide a look into the
past, when such objects were just beginning to form. Tighter
constraints on these sources, both in terms of accurate distance
determinations and deeper optical imaging, should improve our
understanding of the filamentary structures in the vicinity of the
HDF-N and (perhaps) cosmological models. Secure optical counterparts
for many of the CDF-N extended sources are clearly the most important
piece of information currently missing from our picture of these
distant systems. Obtaining further constraints on the X-ray properties
of these sources (e.g., temperatures, abundances, surface density
profiles) will be difficult, given the large amount of observing time
needed simply to detect them. As observations of the CDF-N continue
with both {\it Chandra} and {\it XMM-Newton}, it may be possible to
search for temperature variations between the two peaks in the
tentative high redshift system. This would be a valuable diagnostic
for evaluating the nature of this extended X-ray source (e.g., is the
variation consistent with a merger shocks or cold cluster cores moving
through low density, shock-heated intra-cluster gas).

\acknowledgements

This work would not have been possible without the enormous efforts of
the entire {\it Chandra} and ACIS teams; we particularly thank
P.~Broos, B.~Olsson, and L.~Townsley for data analysis software and
CTI correction support. We thank the referee, H. Ebeling, for
suggestions which improved the final version of this paper. We thank
S.~Dawson and J.~Mulchaey for helpful discussions, A.~Barger,
L.~Cowie, C.~Liu, E.~Richards, and C.~Steidel for kindly providing or
making public their optical and radio images of the HDF-N region,
E.~Bertin and S.~Arnouts for making {\sc SExtractor} available, and
M.~Bolzonella, J.-M.~Miralles and R.~Pell{\' o} for making {\sc
hyperz} available. The HET is a joint project of the University of
Texas at Austin, the Pennsylvania State University, Stanford
University, Ludwig-Maximillians-Universit\"at M\"unchen, and
Georg-August-Universit\"at G\"ottingen. The HET is named in honor of
its principal benefactors, William P. Hobby and Robert E. Eberly. The
Marcario LRS is a joint project of the University of Texas at Austin,
the Instituto de Astronomia de la Universidad Nacional Autonoma de
Mexico, Ludwig-Maximillians-Universit\"at M\"unchen,
Georg-August-Universit\"at G\"ottingen, Stanford University, and the
Pennsylvania State University. We gratefully acknowledge the financial
support of NSF CAREER award AST-9983783 (FEB, DMA, WNB), NASA GSRP
grant NGT~5-50247 and the Pennsylvania Space Grant Consortium (AEH),
NASA grant NAG~5-10875 (TM), NASA grant NAS~8-38252 (GPG, PI), and NSF
grant AST-9900703~(DPS).

%
%

\begin{deluxetable}{lllrrcrrrrrrl}
\rotate
\tabletypesize{\scriptsize}
\tablewidth{0pt}
\tablecaption{Extended X-ray Sources in the CDF-N\label{tab:sources}} 
\tablehead{
\colhead{(1)} & 
\colhead{(2)} & 
\colhead{(3)} & 
\colhead{(4)} & 
\colhead{(5)} & 
\colhead{(6)} & 
\colhead{(7)} & 
\colhead{(8)} & 
\colhead{(9)} & 
\colhead{(10)} & 
\colhead{(11)} & 
\colhead{(12)} &
\colhead{(13)} \\
\colhead{ID} & 
\colhead{CXOHDFN Source} & 
\colhead{Region} & 
\colhead{Soft Counts} & 
\colhead{Full Counts} & 
\colhead{S-to-B Ratio} & 
\colhead{BR} & 
\colhead{$kT$} & 
\colhead{$F_{\rm X}$} & 
\colhead{$L_{\rm X}$} & 
\colhead{$P_{\rm over}$} & 
\colhead{$z$} & 
\colhead{Comments}
}
\startdata
1 & J123557.8$+$621540 & $27''\times27''$             &  76.3 $\pm$ 24.1 &        $<$ 121.5 & 0.18
  & $<$~1.25               & \nodata                 &     5.7 &      4.9? & 1.00\tablenotemark{\dagger} & 0.44?  &     \\
2 & J123620.0$+$621554 & $45''\times25''$, $145\arcdeg$& 273.8 $\pm$ 37.2 & 430.7 $\pm$ 63.6 & 0.32
  & $0.51^{+0.16}_{-0.15}$ & $2.93^{+13.59}_{-1.40}$  &   13.5 &    $>$32.3? & 0.68 & $>$0.68? & No bright optical counterparts   \\
3 & J123645.0$+$621142 & $30''\times30''$             & 100.1 $\pm$ 31.1 &        $<$ 165.0 & 0.16
  & $<$~2.19               & \nodata                 &     3.1 &     20.0 & 1.00 & 1.01\   & Assoc. w. VLA~J123644.3$+$621133 (FR~I)\\
4 & J123704.6$+$621652 & $32''\times32''$             & 121.3 $\pm$ 31.8 & 203.6 $\pm$ 56.2 & 0.17
  & $<$~1.25               & \nodata                 &     6.0 &      3.7? & 0.98 & 0.38?\   &     \\
5 & J123721.2$+$621526 & $22''\times22''$             &  80.3 $\pm$ 22.1 & 108.2 $\pm$ 35.9 & 0.24
  & $<$~1.36               & \nodata                 &     3.6 & 2.5--9.4? & 1.00 & 0.4--0.7?&     \\
6 & J123756.0$+$621506 & $25''\times18''$, $55\arcdeg$& 303.4 $\pm$ 26.3 & 368.5 $\pm$ 38.8 & 0.99
  & $<$~0.28               & $2.93^{+2.80}_{-0.96}$  &    17.1 &      2.1 & 1.00\tablenotemark{\dagger} & 0.19\   &     \\
\enddata
\vspace{-0.5cm} 
\tablecomments{{\it Column~1}: Source number. {\it Column~2}: 
Source name given as CXOHDFN~JHHMMSS.S$+$DDMMSS. {\it Column~3}:
Source extraction region given as major axis and minor axis in
arcseconds, and, if the region is not circular, the position angle in
degrees. {\it Columns 4 and 5}: Background-subtracted 0.5--2.0~keV and
0.5--8.0~keV counts were found with aperture photometry, using the
regions defined in column~3 and individual background annuli as noted
in $\S$\ref{properties}. The standard deviations for the source and
background counts have been computed following the method of
\citet{Gehrels1986} and combined following the ``numerical
method'' described in $\S$1.7.3 of \citet{Lyons1991}. {\it Column~6}:
The ratio of the total number of 0.5--2.0~keV source counts to the
total number of 0.5--2.0~keV background counts expected within the
region defined in column~2. {\it Column~7}: Band ratio (BR),
calculated as the ratio of 2.0--8.0~keV count rate (or 3$\sigma$ upper
limit) to 0.5--2.0~keV count rate. Errors have been combined following
the ``numerical method'' described in $\S$1.7.3 of \citet{Lyons1991}. 
{\it Column~8}: Rest-frame thermal plasma temperature $kT$ as
determined from the best-fit models to the ACIS-I spectra. Also listed
are the 90\% confidence errors calculated for one parameter of
interest ($\Delta\chi^2 = 2.7$). {\it Column~9}: Observed 0.5--2.0~keV
fluxes in units of 10$^{-16}$~erg~cm$^{-2}$~s$^{-1}$ calculated
assuming the best-fit thermal plasma temperature $kT$ listed in
Column~8 or, for sources with too few counts to allow meaningful
spectral fitting, a rest-frame temperature of 1~keV. The fluxes have
been corrected for the portions of the aperture masked out to
eliminate contaminating point sources. {\it Column~10}:
Absorption-corrected rest-frame 0.5--2.0~keV luminosities in units of
10$^{41}$~erg~s$^{-1}$. For sources with too few counts to allow
meaningful spectral fitting, we assumed a neutral hydrogen
column~density of $N_{\rm H} = 1.6\times10^{20}$~cm$^{-2}$. {\it
Column~11}: Monte Carlo simulation probability of an overdensity of
optical galaxies centered on the X-ray emission as compared to field
sources. The probability indicates the fraction of randomly selected
regions with fewer sources than found to be coincident with the X-ray
source. Simulations were performed using either the $I$-band or
$R$-band (denoted by $\dagger$) images; see $\S$~\ref{montecarlo} for
details. {\it Column~12}: Probable redshift (see $\S$\ref{detections}). 
{\it Column~13}: Comments.}
\end{deluxetable}

\begin{deluxetable}{lrrrrrrl}
\tabletypesize{\scriptsize}
\tablewidth{0pt}
\tablecaption{Constraints on Undetected Cluster Candidates in the CDF-N\label{tab:nosources}} 
\tablehead{
\colhead{(1)} & 
\colhead{(2)} & 
\colhead{(3)} & 
\colhead{(4)} & 
\colhead{(5)} & 
\colhead{(6)} & 
\colhead{(7)} \\
\colhead{Source} & 
\colhead{Soft Counts} & 
\colhead{Full Counts} & 
\colhead{$F_{\rm X}$} & 
\colhead{$L_{\rm X}$} & 
\colhead{$z$} & 
\colhead{Comments}
}
\startdata
ClG 1236$+$6215         &  $<$~83.7        & $<$~151.5        &  $<$~4.9 & $<$~21.6 & 0.85\   &  \\
VLA~J123725.7$+$621128  &  $<$~85.9        & $<$~197.6        &  $<$~4.3 & $<$~26.4--142.5 & $\sim 1-2$? & Wide angle tail radio source \\
\enddata
\vspace{-0.5cm} 
\tablecomments{{\it Column~1}: Sources of potential extended X-ray emission 
which are not formally detected (see $\S$\ref{nondetections}); ClG
1236$+$6215 is shown in Figure~\ref{fig:clusterover}, since it appears
to have some associated point-source emission. {\it Columns 2 and 3}:
Upper limits on the 0.5--2.0~keV and 0.5--8.0~keV counts were found
with aperture photometry, using 30$\arcsec$ radius apertures as noted
in $\S$\ref{detections}. The upper limits are calculated at the
3$\sigma$ level for Gaussian statistics. {\it Column~4}: Observed
0.5--2.0~keV flux limits in units of 10$^{-16}$~erg~cm$^{-2}$~s$^{-1}$
calculated assuming an absorbed MEKAL single temperature thermal
plasma spectrum with $N_{\rm H} = 1.6\times10^{20}$ cm$^{-2}$ and
$kT=1.0$~keV. Large deviations from the assumed temperature should not change
the flux by more than $\sim$20\%. The fluxes have been corrected for
the portions of the aperture masked out to eliminate contaminating
point sources. {\it Column~5}: Absorption-corrected 0.5--2.0~keV
luminosity limits in units of 10$^{41}$~erg~s$^{-1}$. {\it Column~6}:
Probable redshift (see $\S$\ref{nondetections}). {\it Column~7}: Comments.}
\end{deluxetable}

\clearpage

%
%

\appendix

\section{Appendix A: Hobby-Eberly Telescope Observations}\label{app:HET}

Moderate resolution long-slit spectroscopic observations were taken
for two galaxies (123755.7$+$621507 and 123756.1$+$621514) that are
likely to be associated with source~6. The spectra were taken during
the moonless nights of 2001 June 13 and 14 with the LRS mounted at the
prime focus of the HET. A 300 line mm$^{-1}$ grism blazed at 5500~\AA\
was used with a GG-385 UV-blocking filter. The detector was a thinned,
antireflection-coated $3072\times1024$ pixel$^{2}$ Ford Aerospace CCD
and was binned $2\times2$ during readout; this produced an image scale
of $0\farcs50$ pixel$^{-1}$ and a dispersion of 4.5~\AA~pixel$^{-1}$. 
The spectra covered the range from 4150--9000~\AA\ at a resolution of
$\approx$~20~\AA. A $2\arcsec$ slit was aligned at a position angle of
$15\arcdeg$ during the first observation and exposed for 1300~s in an
attempt to obtain simultaneous spectra of both objects. Only
123756.1$+$621514 was exposed with enough signal-to-noise to extract a
useful spectrum. A second observation of 1350~s was taken the next
evening, this time with the slit aligned along the disk of
123755.7$+$621507 (i.e., at a slit position angle of $-20\arcdeg$) to
increase the signal-to-noise ratio of the spectrum. Both observations
were taken close to transit, with airmasses of 1.25 and 1.24,
respectively. Conditions were less than ideal on the first night but
were generally transparent and photometric on the second night; note
that the first observation was performed shortly before the telescope
closed down due to dust and high winds, and it appears that the
spectrum of 123756.1$+$621514 is consequently contaminated by dust
extinction longward of $\sim$6000~\AA. Unfortunately a flux standard
star was not obtained during the first night, and thus we have used
the flux standard from the second night to calibrate crudely the
spectrum of 123756.1$+$621514. Since we care only about the absorption
features in the spectrum to extract a redshift, the overall shape of
the continuum and its absolute normalization are not important. The
flux standard star BD $+17^\circ4708$ \citep{Oke1983, Massey1988} was
observed during the second night at the same slit width and comparable
air mass to flux calibrate our spectra. Ne and HgCdZn lamp exposures
were taken each night to provide wavelength calibration lines. The
wavelength calibration was confirmed using several narrow night sky
emission lines (e.g., [\ion{O}{1}] $\lambda$5577, $\lambda$6300) and
should be accurate to $\sim$1~\AA. At the beginning of each night, a
series of quartz lamp spectra for flat fielding were obtained as well
as a series of bias frames to remove residual structure in the DC
offset not accounted for by the overscan region. No attempt has been
made to remove the dark current, as it should be negligible. 

The spectra were reduced using standard {\sc iraf} procedures, and
redshifts were assessed using the {\sc rv} package and a template
spectrum of the E0 elliptical galaxy NGC 3379 obtained at comparable
spectral resolution. The key spectral features used to assess the
redshifts of these two galaxies were the
\ion{Ca}{2} band, CH~G band, and \ion{Mg}{1}~b absorption lines.
Both 123755.7$+$621507 and 123756.1$+$621514 were found to lie at
redshifts of $z=0.190$. The flux-calibrated, extinction-corrected,
rest-frame spectra for both sources are shown in
Figure~\ref{fig:optspectra}. To assess the accuracy of the redshift
measurements and confirm the optical morphologies seen in
Figure~\ref{fig:overlay}, we also display two spectral models from the
PEGASE spectral synthesis templates
\citep{Fioc97} overlaid on the HET spectra. 123755.7$+$621507, which
appears to be an elliptical or bulge-dominated galaxy, is best fit
with an old stellar population template. 123756.1$+$621514, which
appears to be a edge-on disk galaxy, is best fit with a young stellar
population template. While these templates are not perfect matches to
the data, they do clearly highlight the stellar absorption features of
the respective galaxies. Note that the drop in the spectrum of
123756.1$+$621514 above 6000~\AA\ is likely due to atmospheric dust
extinction.

\setcounter{figure}{0}
\begin{figure*}
\vspace{-0.1in}
\centerline{
\includegraphics[width=9cm]{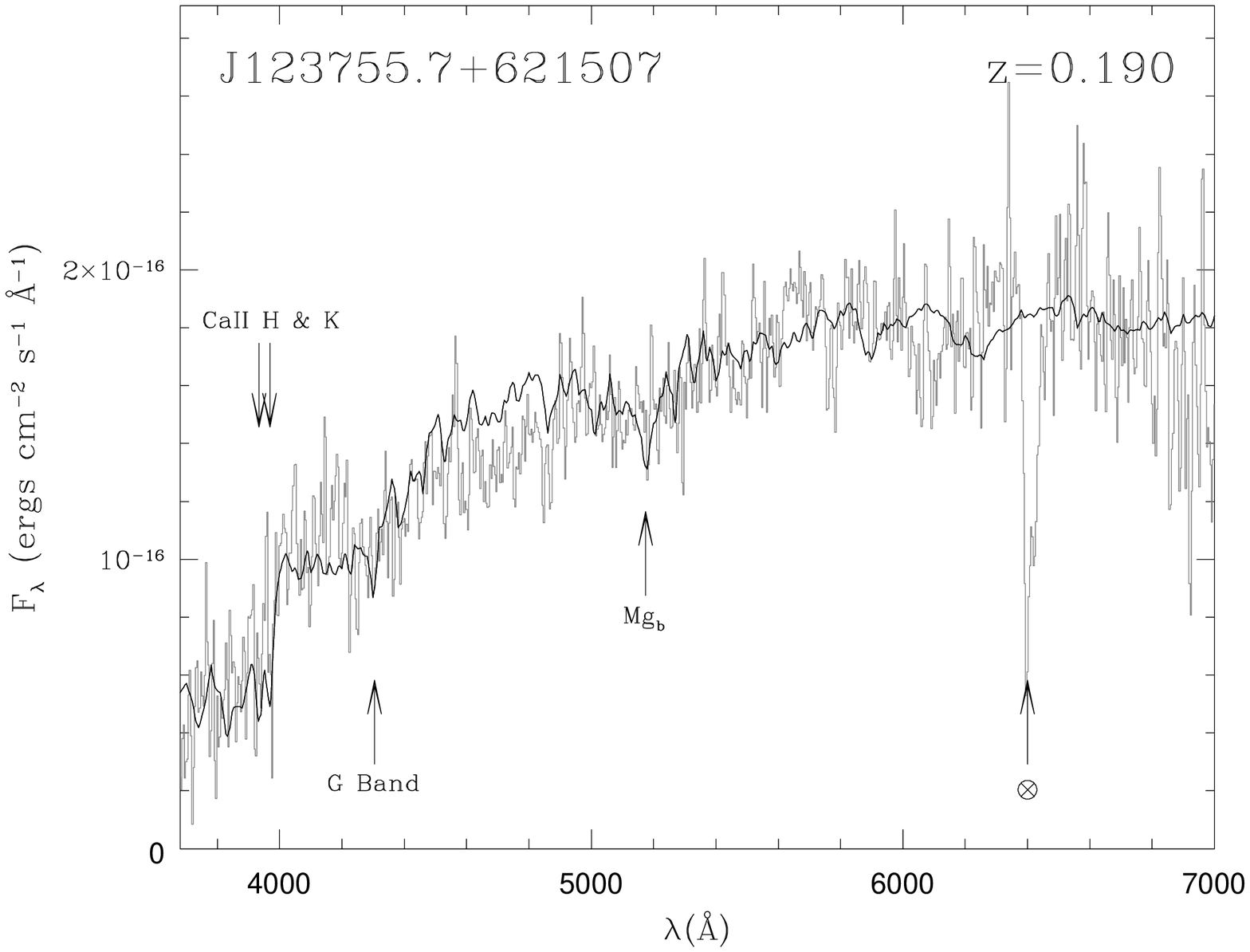}\hfill
\includegraphics[width=9cm]{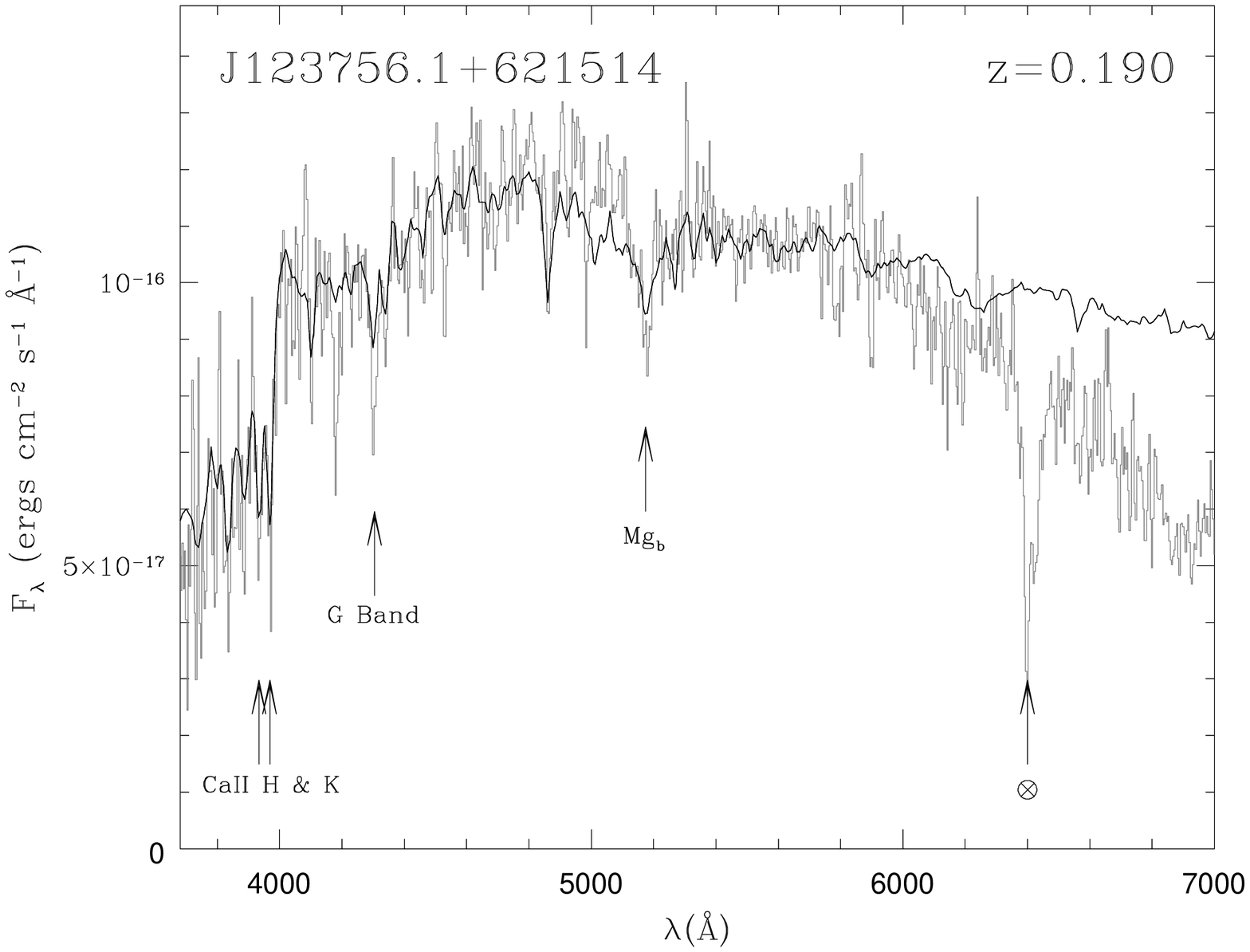} 
} 
\vspace{-1.6cm} 
\figcaption[fig7.eps]{HET spectra (gray) of two bright sources,
123755.7$+$621507 and 123756.1$+$621514, which coincide with the
center of the source~6. The spectra have been flux-calibrated,
extinction-corrected, and shifted to rest-frame wavelengths. Both
sources are found to lie at $z=0.190$. The strong absorption feature
at $\sim$6400~\AA\ is due to the atmospheric oxygen B band
absorption. Note that the drop in the spectrum of 123756.1$+$621514
above 6000~\AA\ is unlikely to be intrinsic and is probably due to
atmospheric dust extinction. Overlaid on the HET spectra are two
spectral models (black) from the PEGASE spectral synthesis templates
\citep{Fioc97}, one of an old stellar population (e.g., an elliptical
galaxy) for 123755.7$+$621507 and one of a young stellar population
(e.g., a disk-dominated spiral) for 123756.1$+$621514.
\label{fig:optspectra}}
\end{figure*}


\begin{thebibliography}{}

\bibitem[Alexander et al.(2001)]{Alexander2001} Alexander, D.~M., 
Brandt, W.~N., Hornschemeier, A.~E., Garmire, G.~P., Schneider, D.~P.,
Bauer, F.~E., \& Griffiths, R.~E.\ 2001, \aj, 122, 2156

\bibitem[Alexander et al.(2002)]{Alexander2002} Alexander, D.~M., 
Vignali, C., Bauer, F.~E., Brandt, W.~N., Hornschemeier, A.~E.,
Garmire, G.~P., \& Schneider, D.~P. \ 2002, \aj, in press
(astro-ph/0111397)

\bibitem[Allen \& Fabian(1998)]{Allen1998} Allen, S.~W.~\& 
Fabian, A.~C.\ 1998, \mnras, 297, L57 

\bibitem[Arnaud(1996)]{Arnaud1996} Arnaud, K.~A. 1996, in {ASP} 
Conf. Ser. 101, Astronomical Data Analysis Software and Systems V, ed. 
G.~Jacoby, \& J.~Barnes (San Francisco: ASP), 17

\bibitem[Barger, Cowie, \& Richards(2000)]{Barger2000} Barger, 
A.~J., Cowie, L.~L., \& Richards, E.~A.\ 2000, \aj, 119, 2092 

\bibitem[Barger et al.(1999)]{Barger1999} Barger, A.\ J., Cowie, 
L.\ L., Trentham, N., Fulton, E., Hu, E.\ M., Songaila, A., \& Hall, D.\ 
1999, \aj, 117, 102 

\bibitem[Bertin \& Arnouts(1996)]{Bertin1996} Bertin, E.\ \& 
Arnouts, S.\ 1996, \aaps, 117, 393 

\bibitem[Blanton et al.(2001)]{Blanton2001} Blanton, E.~L., Gregg, 
M.~D., Helfand, D.~J., Becker, R.~H., \& Leighly, K.~M.\ 2001, \aj, 121, 
2915 

\bibitem[Bolzonella, Miralles, \& Pell{\' o}(2000)]{Bolzonella2000} 
Bolzonella, M., Miralles, J., \& Pell{\' o}, R.\ 2000, \aap, 363, 476 

\bibitem[Bond, Kofman, \& Pogosyan(1996)]{Bond1996} 
Bond, J.~R., Kofman, L., \& Pogosyan, D.\ 1996, \nat, 380, 603

\bibitem[Brandt et al.(2001a)]{Brandt2001a} Brandt, W.\ N.\ et al.\ 
2001a, \aj, 122, 1 (Paper~IV)

\bibitem[Brandt et al.(2001b)]{Brandt2001b} Brandt, W.\ N., et al.\ 2001b, 
\aj, in press (astro-ph/0108404; Paper~V)

\bibitem[Brinkmann et al.(1999)]{Brinkmann1999} Brinkmann, W., 
Chester, M., Kollgaard, R., Feigelson, E., Voges, W., \& Hertz, P.\ 1999, \aaps, 134, 221

\bibitem[Bruzual \& Charlot(1993)]{Bruzual1993} Bruzual A., 
G.~\& Charlot, S.\ 1993, \apj, 405, 538 

\bibitem[Burg et al.(1992)]{Burg1992} Burg, R.~et al.\ 1992, 
\aap, 259, L9 

\bibitem[Burns et al.(1994)]{Burns1994} Burns, 
J.\ O., Rhee, G., Owen, F.\ N., \& Pinkney, J.\ 1994, \apj, 423, 94 

\bibitem[Burns et al.(1996)]{Burns1996} Burns, J.~O., Ledlow, 
M.~J., Loken, C., Klypin, A., Voges, W., Bryan, G.~L., Norman, M.~L., \& 
White, R.~A.\ 1996, \apjl, 467, L49 

\bibitem[Cavaliere, Giacconi, \& Menci(2000)]{Cavaliere2000} 
Cavaliere, A., Giacconi, R., \& Menci, N.\ 2000, \apjl, 528, L77

\bibitem[Cavaliere, Menci, \& Tozzi(1997)]{Cavaliere1997} Cavaliere, 
A., Menci, N., \& Tozzi, P.\ 1997, \apjl, 484, L21

\bibitem[Cen \& Ostriker(1999)]{Cen1999} Cen, R.~\& Ostriker, 
J.~P.\ 1999, \apj, 514, 1 

\bibitem[Chapman, McCarthy, \& Persson(2000)]{Chapman2000} Chapman, 
S.~C., McCarthy, P.~J., \& Persson, S.~E.\ 2000, \aj, 120, 1612 

\bibitem[Cohen et al.(2000)]{Cohen2000} Cohen, J.\ G., Hogg, D.\ 
W., Blandford, R., Cowie, L.\ L., Hu, E., Songaila, A., Shopbell, P., \& 
Richberg, K.\ 2000, \apj, 538, 29 

\bibitem[Coleman, Wu, \& Weedman(1980)]{Coleman1980} Coleman, 
G.~D., Wu, C.-C., \& Weedman, D.~W.\ 1980, \apjs, 43, 393 

\bibitem[Cowie et al.(2002)]{Cowie2002}
Cowie L.\ L., Garmire G.\ P., Bautz M.\ W., Barger A.\ J., Brandt W.\ N., \&
Hornschemeier A.\ E. 2002, ApJ, submitted

\bibitem[Dawson et al.(2001)]{Dawson2001} Dawson, S., Stern, D., Bunker, A.\ J., Spinrad, H., 
\& Dey, A.\ 2001, \aj, 122, 598

\bibitem[Donahue et al.(2001)]{Donahue2001} Donahue, M.~et al.\ 
2001, \apjl, 552, L93 

\bibitem[Dobrzycki et al.(1999)]{Dobrzycki1999} 
Dobrzycki, A., Ebeling, H., Glotfelty, K., Freeman, P., Damiani, F.,
Elvis, M., \& Calderwood, T. 1999, {\it Chandra} Detect 1.0 User
Guide. {\it Chandra} X-ray Center, Cambridge

\bibitem[Ebeling et al.(1997)]{Ebeling1997} Ebeling, H., Edge, 
A.~C., Fabian, A.~C., Allen, S.~W., Crawford, C.~S., \& Boehringer, H.\ 
1997, \apjl, 479, L101 

\bibitem[Ebeling, White, \& Rangarajan(2001)]{Ebeling2001}
Ebeling, H., White, D.\ A., \& Rangarajan, F.\ V.\ N.\ 2001, \mnras, submitted

\bibitem[Ebeling \& Wiedenmann(1993)]{Ebeling1993} Ebeling, H.~\& 
Wiedenmann, G.\ 1993, \pre, 47, 704 

\bibitem[Fanaroff \& Riley(1974)]{Fanaroff1974} Fanaroff, B.\ L.\ \& 
Riley, J.\ M.\ 1974, \mnras, 167, 31P 

\bibitem[Ferguson, Dickinson, \& Williams(2000)]{Ferguson2000} 
Ferguson, H.\ C., Dickinson, M., \& Williams, R.\ 2000, \araa, 38, 667 

\bibitem[Fern{\' a}ndez-Soto et al.(1999)]{Fernandez1999} Fern{\' a}ndez-Soto, A., Lanzetta, K.\ 
M., \& Yahil, A.\ 1999, \apj, 513, 34 

\bibitem[Fioc \& Rocca-Volmerange(1997)]{Fioc97} Fioc, M. 
\& Rocca-Volmerange, B. 1997, \aap, 326, 950 

\bibitem[Gehrels(1986)]{Gehrels1986} Gehrels, N.\ 1986, \apj, 303, 
336 

\bibitem[Gladders \& Yee(2000)]{Gladders2000} Gladders, M.~D.~\& Yee, H.~K.~C.\ 2000, \aj, 120, 2148

\bibitem[Gomez et al.(1997)]{Gomez1997} Gomez, P.\ L., Pinkney, 
J., Burns, J.\ O., Wang, Q., Owen, F.\ N., \& Voges, W.\ 1997, \apj, 474, 
580 

\bibitem[Hasinger et al.(1998)]{Hasinger1998} Hasinger, G.~et al.\ 
1998, \aap, 340, L27 

\bibitem[Helsdon \& Ponman(2000)]{Helsdon2000} Helsdon, S.\ F.\ \& 
Ponman, T.\ J.\ 2000, \mnras, 319, 933 

\bibitem[Hill(2000)]{Hill2000} Hill, G.~J.\ 2000, \procspie, 
4008, 50 

\bibitem[Hill et al.(1998)]{Hill1998} Hill, G.~J., Nicklas, 
H.~E., MacQueen, P.~J., Tejada, C., Cobos Duenas, F.~J., \& Mitsch, W.\ 
1998, \procspie, 3355, 375 

\bibitem[Hoessel \& Schneider(1985)]{Hoessel1985} Hoessel,
J.~G.~\& Schneider, D.~P.\ 1985, \aj, 90, 1648

\bibitem[Hornschemeier et al.(2001)]{Hornschemeier2001} Hornschemeier, 
A.\ E.\ et al.\ 2001, \apj, 554, 742

\bibitem[Kaastra(1992)]{Kaastra1992} Kaastra, J.S. 1992, 
An X-Ray Spectral Code for Optically Thin Plasmas
(Internal SRON-Leiden Report, updated version 2.0)

\bibitem[Kuntz(2000)]{Kuntz2000} Kuntz, K.\ D.\ 2000,
Ph.D.\ Thesis, University of Maryland, 147 

\bibitem[Lehmann et al.(2000)]{Lehmann2000} Lehmann, I.~et al.\ 
2000, From Extrasolar Planets to Cosmology: The VLT Opening Symposium, 
Proceedings of the ESO Symposium held at Antofagasta, Chile, 1-4 March 
1999.~Edited by Jacqueline Bergeron and Alvio Renzini.~Berlin: 
Springer-Verlag, 2000.~p.~121.

\bibitem[Lehmann et al.(2001)]{Lehmann2001} Lehmann, I.~et al.\ 
2001, \aap, 371, 833 

\bibitem[Liedahl, Osterheld, \& Goldstein(1995)]{Liedahl1995} 
Liedahl, D.~A., Osterheld, A.~L., \& Goldstein, W.~H.\ 1995, \apjl, 438, 
L115 

\bibitem[Liu et al.(1999)]{Liu1999} Liu, C.\ 
T., Petry, C.\ E., Impey, C.\ D., \& Foltz, C.\ B.\ 1999, \aj, 118, 1912 

\bibitem[Lyons(1991)]{Lyons1991} Lyons, L. 1991, Data Analysis for
Physical Science Students. (Cambridge: Cambridge University Press)

\bibitem[Massey et al.(1988)]{Massey1988} Massey, P., Strobel, K., 
Barnes, J. V., \&, Anderson, E. 1988, \apj, 328, 315

\bibitem[McHardy et al.(1998)]{McHardy1998} McHardy, I.~M.~et al.\ 
1998, \mnras, 295, 641 

\bibitem[Mewe, Gronenschild, \& van den Oord(1985)]{Mewe1985} 
Mewe, R., Gronenschild, E.~H.~B.~M., \& van den Oord, G.~H.~J.\ 1985, 
\aaps, 62, 197 

\bibitem[Metzler \& Evrard(1994)]{Metzler1994} Metzler, C.~A.~\& 
Evrard, A.~E.\ 1994, \apj, 437, 564 

\bibitem[Mohr \& Evrard(1997)]{Mohr1997} Mohr, J.~J.~\& Evrard, 
A.~E.\ 1997, \apj, 491, 38 

\bibitem[Mukai(2000)]{Mukai2000} Mukai, K. 2000, {\sc pimms} Version 3.0 Users' Guide. NASA/GSFC, Greenbelt

\bibitem[Mulchaey(2000)]{Mulchaey2000} Mulchaey, J.\ S.\ 2000, 
\araa, 38, 289 

\bibitem[Mulchaey et al.(1996)]{Mulchaey1996} Mulchaey, J.\ S., Davis, D.\ S., 
Mushotzky, R.\ F., \& Burstein, D.\ 1996, \apj, 456, 80 

\bibitem[Mullis et al.(2001)]{Mullis2001} Mullis, C.~R., Henry, 
J.~P., Gioia, I.~M., B{\" o}hringer, H., Briel, U.~G., Voges, W., \& 
Huchra, J.~P.\ 2001, \apjl, 553, L115 

\bibitem[Muxlow et al.(1999)]{Muxlow1999} Muxlow, T.~W.~B., 
Wilkinson, P.~N., Richards, A.~M.~S., Kellermann, K.~I., Richards, E.~A., 
\& Garrett, M.~A.\ 1999, New Astronomy Review, 43, 623 

\bibitem[Oke \& Gunn(1983)]{Oke1983} Oke, J.~B.~\& Gunn, J.~E.\ 
1983, \apj, 266, 713 

\bibitem[Oukbir, Bartlett, \& Blanchard(1997)]{Oukbir1997} Oukbir, 
J., Bartlett, J.~G., \& Blanchard, A.\ 1997, \aap, 320, 365 

\bibitem[Pierre, Bryan, \& Gastaud(2000)]{Pierre2000} Pierre, M., 
Bryan, G., \& Gastaud, R.\ 2000, \aap, 356, 403 

\bibitem[Pildis, Evrard, \& Bregman(1996)]{Pildis1996} Pildis, 
R.~A., Evrard, A.~E., \& Bregman, J.~N.\ 1996, \aj, 112, 378 

\bibitem[Poggianti(1997)]{Poggianti1997} Poggianti, B.~M.\ 1997,
\aaps, 122, 399

\bibitem[Ponman et al.(1996)]{Ponman1996} 
Ponman, T.\ J., Bourner, P.\ D.\ J., Ebeling, H., \& Bohringer, H.\ 1996, 
\mnras, 283, 690 

\bibitem[Postman \& Lauer(1995)]{Postman1995} Postman, M.~\& Lauer, 
T.~R.\ 1995, \apj, 440, 28 

\bibitem[Raymond \& Smith(1977)]{Raymond1977} Raymond, J.\ C., \& 
Smith, B.\ W.\ 1977, \apjs, 35, 419 

\bibitem[Richards(1999)]{Richards1999a} Richards, E.\ A.\ 1999, 
Ph.D.\ Thesis, University of Virginia, 171 (see also \pasp\  112, 1001)

\bibitem[Richards et al.(1999)]{Richards1999b} Richards, E.\ A., 
Fomalont, E.\ B., Kellermann, K.\ I., Windhorst, R.\ A., Partridge, R.\ B., 
Cowie, L.\ L., \& Barger, A.\ J.\ 1999, \apjl, 526, L73 

\bibitem[Richards et al.(1998)]{Richards1998} Richards, E.\ A., 
Kellermann, K.\ I., Fomalont, E.\ B., Windhorst, R.\ A., \& Partridge, R.\ 
B.\ 1998, \aj, 116, 1039 

\bibitem[Ricker \& Sarazin(2001)]{Ricker2001} Ricker, P.~M.~\& Sarazin, C.~L.\ 2001, \apj, 561, 621

\bibitem[Rosati et al.(1998)]{Rosati1998} Rosati, P., della Ceca, R., Norman, 
C., \& Giacconi, R.\ 1998, \apjl, 492, L21 

\bibitem[Rudnick \& Owen(1976)]{Rudnick1976} Rudnick, L.\ \& Owen, 
F.\ N.\ 1976, \apjl, 203, L107 

\bibitem[Smith et al.(2001)]{Smith2001} Smith, G.~P.\ et\ al.\ 2001, \mnras, in press (astro-ph/0109465)

\bibitem[Snellen \& Best(2001)]{Snellen2001} Snellen, I.~A.~G.~\& Best, P.~N.\ 2001, \mnras, 328, 897

\bibitem[Snowden et al.(1998)]{Snowden1998} Snowden, S.~L., Egger, 
R., Finkbeiner, D.~P., Freyberg, M.~J., \& Plucinsky, P.~P.\ 1998, \apj, 
493, 715 

\bibitem[Stark et al.(1992)]{Stark1992} Stark, A.\ A., Gammie, 
C.\ F., Wilson, R.\ W., Bally, J., Linke, R.\ A., Heiles, C., \& Hurwitz, 
M.\ 1992, \apjs, 79, 77 

\bibitem[Steidel \& Hamilton(1993)]{Steidel1993} Steidel, C.\ C.\ 
\& Hamilton, D.\ 1993, \aj, 105, 2017 

\bibitem[Townsley et al.(2000)]{Townsley2000} 
Townsley, L.~K., Broos, P.~S., Garmire, G.~P., \& Nousek, J.~A.\ 2000, 
\apjl, 534, L139 

\bibitem[Tully(1987)]{Tully1987} Tully, R.~B.\ 1987, \apj, 323, 1 

\bibitem[Vikhlinin et al.(1998)]{Vikhlinin1998} Vikhlinin, A., 
McNamara, B.~R., Forman, W., Jones, C., Quintana, H., \& Hornstrup, A.\ 
1998, \apj, 502, 558 

\bibitem[Weisskopf et al.(2000)]{Weisskopf2000} Weisskopf, M.\ C., 
Tananbaum, H.\ D., Van Speybroeck, L.\ P., \& O'Dell, S.\ L. \ 2000,
\procspie, 4012, 2

\bibitem[Williams et al.(1996)]{Williams1996} Williams, R.\ E.\ et 
al.\ 1996, \aj, 112, 1335 

\bibitem[Xue \& Wu(2000)]{Xue2000} Xue, Y.\ \& Wu, X.\ 2000, 
\apj, 538, 65 

\bibitem[Zamorani et al.(1999)]{Zamorani1999} Zamorani, G.~et al.\ 
1999, \aap, 346, 731 

\end{thebibliography}
\end{document}